\documentclass[12pt]{article}
\pdfoutput=1
\usepackage{hyperref}
\usepackage{textcomp}
\usepackage[DIV13]{typearea}
\usepackage{amsmath,amssymb,latexsym,dsfont}
\usepackage{slashed, simplewick}
\usepackage[utf8]{inputenc}
\usepackage{graphicx,float,placeins}
\usepackage{makeidx}
\usepackage[font=small,labelfont=bf]{caption}
\usepackage{nicefrac}
\usepackage{subfigure}
\usepackage{array, bigdelim,multirow,multicol}
\usepackage[usenames]{color}
\usepackage{soul,appendix}
\graphicspath{{immagini/},{plots/}}
\def\cA{{\cal A}} \def\cB{{\cal B}}

\def\cL{{\cal L}} 
 \def\cO{{\cal O}}

\newcommand{\TeV}{\;\text{TeV}}

\newcommand{ \mysmall}[1]{\scriptscriptstyle #1}
\newcommand{\hc}{\text{h.c.}}
\newcommand{\ba} {\begin{eqnarray}}
\newcommand{\ea} {\end{eqnarray}}
\newcommand{\nn}{\nonumber}
\newcommand{\be}{\begin{equation}}
\newcommand{\ee}{\end{equation}}
\newcommand{\bea}{\begin{eqnarray}}
\newcommand{\eea}{\end{eqnarray}}

\newcommand{\vir}{\; ,}

\newcommand{\arXhref}[1]{\href{http://arxiv.org/abs/#1}{#1}}

\newcommand{\RK} {R^{\mu/e}_K}
\newcommand{\RKs} {R^{\mu/e}_{K^\ast}}
\newcommand{\RKss} {R^{\mu/e}_{K^{(\ast)}}}
\newcommand{\RD} {R^{\tau/\ell}_{D^{(*)}}}

\allowdisplaybreaks
\begin{document}
 \unitlength = 1mm

\setlength{\extrarowheight}{0.2 cm}

\thispagestyle{empty}

\bigskip

\vskip 1cm

\begin{center}
\vspace{1.5cm}
    {\Large\bf On the Importance of Electroweak Corrections\\ [0.5 cm] for ${\mathbf B}$ Anomalies} \\[1cm]
   {\bf Ferruccio Feruglio$^{a}$, Paride Paradisi$^{a}$, Andrea Pattori$^b$}    \\[0.5cm]
     {\em $(a)$  Dipartimento di Fisica e Astronomia `G.~Galilei', Universit\`a di Padova\\
INFN, Sezione di Padova, Via Marzolo~8, I-35131 Padua, Italy}  \\
  {\em $(b)$  Physik-Institut, Universit\"at Z\"urich, CH-8057 Z\"urich, Switzerland} \\[1.0cm]
\end{center}

\centerline{\large\bf Abstract}
\begin{quote}
\indent
The growing experimental indication of Lepton Flavour Universality Violation (LFUV) both in charged- and 
neutral-current semileptonic B-decays, has triggered many theoretical interpretations of such non-standard phenomena. 
Focusing on popular scenarios where the explanation of these anomalies requires New Physics at the TeV scale,
we emphasise the importance of including electroweak corrections to obtain trustable predictions for the models in question.
We find that the most important quantum effects are the modifications of the leptonic couplings of the $W$ and $Z$ 
vector bosons and the generation of a purely leptonic effective Lagrangian. 
Although our results do not provide an inescapable no-go theorem for the explanation of the B anomalies, the tight experimental bounds 
on Z-pole observables and $\tau$ decays challenge an explanation of the current non-standard data.
We illustrate how these effects arise, by providing a detailed discussion of the running and matching procedure which is necessary to derive the 
low-energy effective Lagrangian.
\end{quote}

\newpage 

\section{Introduction}
The search for lepton flavour universality violation (LFUV) represents one of the most powerful tool to 
unveil New Physics (NP) phenomena, as the Standard Model (SM) predicts negligible LFUV effects.
Interestingly enough, in the last few years, hints of large LFUV in semi-leptonic $B$ decays were 
observed by various experimental collaborations both in charged-current as well as neutral-current transitions. 
In particular, the statistically most significant results are accounted for by the following observables:
\begin{align}
R^{\tau/\ell}_{D^*} &=\frac{ \cB(B \to D^* \tau \overline{\nu})_{\rm exp}/\cB(B \to D^* \tau \overline{\nu})_{\rm SM} }{ \cB(B \to D^* \ell \overline{\nu} )_{\rm exp}/ \cB(B \to D^* \ell \overline{\nu} )_{\rm SM} } = 
 1.23 \pm 0.07~, \label{eq:RDexp}  \\
R^{\tau/\ell}_{D } &= \frac{ \cB(B \to D  \tau \overline{\nu})_{\rm exp}/\cB(B \to D  \tau \overline{\nu})_{\rm SM} }{ \cB(B \to D  \ell \overline{\nu} )_{\rm exp}/ \cB(B \to D  \ell \overline{\nu} )_{\rm SM} } =   1.34 \pm 0.17~, \label{eq:RDSexp}
\end{align}
where $\ell=e, \mu$, which follow from the HFAG averages~\cite{Amhis:2016xyh} of Babar~\cite{Lees:2013uzd}, Belle~\cite{Hirose:2016wfn}, 
and LHCb data~\cite{Aaij:2015yra}, combined with the corresponding theory predictions~\cite{Fajfer:2012vx,Aoki:2016frl}, and
\begin{align}
\RKs &=  \left. \frac{ \cB(B \to K^* \mu \bar{\mu})_{\rm exp} }{ \cB(B \to K^* e \bar{e} )_{\rm exp} } \right|_{q^2\in[1.1,6]{\rm GeV}} =  0. 685 {}^{+0.113}_{-0.069} \pm 0.047~,
\label{eq:RKSexp} \\
\RK &=  \left. \frac{ \cB(B \to K \mu \bar{\mu})_{\rm exp} }{ \cB(B \to K e \bar{e} )_{\rm exp} } \right|_{q^2\in[1,6]{\rm GeV}} =  0. 745 {}^{+0.090}_{-0.074} \pm 0.036~,
\label{eq:RKexp}
\end{align}
based on combination of LHCb data~\cite{Aaij:2014ora,Bifani:2017} with the SM expectation $\RKss =1.00 \pm 0.01$~\cite{Bordone:2016gaq}.
Moreover, there are additional tensions between the SM predictions and experimental data in $b\to s \ell\overline{\ell}$ differential observables,
though large non-perturbative effects can be invoked to explain the observed anomaly~\cite{Jager:2012uw}. 
Yet, it is interesting that the whole set of $b\to s \ell\overline{\ell}$ data could be reconciled with the theory predictions assuming some NP 
contributions exclusively in the muonic channels, see e.g.~Ref.~\cite{Hiller:2014yaa}.
In the recent literature, many studies focused on the experimental signatures implied by the solution of these anomalies in specific scenarios, 
including kaon observables~\cite{Crivellin:2016vjc}, kinematic distributions in $B$ decays~\cite{Becirevic:2016zri}, the lifetime 
of the $B_c^-$ meson~\cite{Alonso:2016oyd}, $\Upsilon$ and $\psi$ leptonic decays~\cite{Aloni:2017eny},
tau lepton searches~\cite{Buttazzo:2016kid} and dark matter~\cite{Sierra:2015fma}.

These anomalies have also triggered many theoretical speculations about the possible NP scenarios at work.
Of particular interest are those attempting to a simultaneous explanation of both charged- and neutral-current anomalies. 
Such a task can be most naturally achieved assuming that NP intervenes through effective 4-fermion operators involving 
left-handed currents, $(\bar{s}_L\gamma_\mu b_L)(\bar{\mu}_L\gamma_\mu \mu_L)$ and
$(\bar{c}_L\gamma_\mu b_L)(\bar{\tau}_L\gamma_\mu \nu_L)$, which are related by the $SU(2)_L$ gauge symmetry~\cite{Bhattacharya:2014wla}. 
In this setup, a necessary requirement is that NP couples much more strongly to the third generation than to the first two, 
since $(\bar{c}_L\gamma_\mu b_L)(\bar{\tau}_L\gamma_\mu \nu_L)$ is already generated at the tree level in the SM while
$(\bar{s}_L\gamma_\mu b_L)(\bar{\mu}_L\gamma_\mu \mu_L)$ is loop-induced.
Such a requirement is automatically accomplished if NP is coupled, in the interaction basis, only to the third 
fermion generation, couplings to lighter generations being generated by the misalignment between the mass and the 
interaction bases through small flavour mixing angles~\cite{Glashow:2014iga}. In this case LFUV is expected to be associated 
with lepton flavour violating (LFV) phenomena. Another possibility consists in NP coupling to different fermion generations 
proportionally to the charged lepton mass squared~\cite{Alonso:2015sja}. In this case LFUV does not necessarily imply 
LFV at an observable level, if also NP preserves the lepton family numbers in the limit of massless neutrinos.

In Ref.~\cite{Feruglio:2016gvd}, electroweak corrections for B anomalies has been analyzed, focusing on a class of 
semileptonic operators defined above the electroweak scale $v$ which are invariant under the full SM gauge group, along the lines of
Refs.~\cite{Bhattacharya:2014wla,Glashow:2014iga,Alonso:2015sja,Fajfer:2012jt,Alonso:2014csa,Buras:2014fpa,Calibbi:2015kma}. 
The main new development of Ref.~\cite{Feruglio:2016gvd} compared to previous studies was the construction of the low-energy effective 
Lagrangian taking into account the running of the Wilson coefficients of a suitable operator basis~\cite{Grzadkowski:2010es} and the 
matching conditions when mass thresholds are crossed. 
The new quantum effects pointed out in Ref.~\cite{Feruglio:2016gvd} do not represent just a correction to the leading order results 
commonly employed in the literature, as one would naively expect. Indeed, the low-energy effective Lagrangian 
contains terms that are absent at the tree-level. Such new terms are crucial in order to establish the predictions of the model in question.
At the quantum level, the leptonic couplings of the $W$ and $Z$ vector bosons are modified and a purely leptonic effective Lagrangian
is also generated. The resulting LFUV in $Z$ and $\tau$ decays, which is correlated with the B-anomalies, and $\tau$ LFV 
contributions turned out to be large, challenging an explanation of such anomalies. Such a conclusion applies under the assumptions
and approximations that will be clarified in this paper and should not be taken as a no-go theorem for the explanation of the B anomalies.
We rather think that the main point raised by our analysis is that including electroweak corrections is mandatory when addressing the
experimental anomalies with new physics at the TeV scale.

Aim of the present work is to detail, complete and expand the results of Ref.~\cite{Feruglio:2016gvd}.
First of all we will derive the full effective Lagrangian relevant to leptonic and semileptonic transitions 
both at the electroweak scale and at the $\tau$-mass scale. After discussing our starting assumptions,
in particular the operators dominating NP effects at the TeV scale, we present the minimal set of 
$SU(2)\times U(1)$ gauge-invariant operators involved in the renormalization group equations (RGE) flow from the TeV to the electroweak scale.
We solve the one-loop RGE equation in the limit 
of exact electroweak symmetry~\cite{Jenkins:2013wua,Alonso:2013hga} and in the leading logarithmic approximation. 
We analyze the induced modification to the
$Z$ couplings, relevant to precision tests. We explicitly show how the scale dependence 
of the RGE contributions from gauge and top Yukawa interactions cancels with that of the matrix elements in the
physical amplitude for the Z decay into a lepton pair. Then we analyze the effective theory below
the electroweak scale by explicitly discussing the matching to an electromagnetic invariant effective
Lagrangian, after integrating out the top quark and the $W$ and $Z$ bosons. Finally we include the
further running, dominated by pure electromagnetic effects, down to the tau lepton mass scale,
after crossing the bottom threshold. 
This discussion is detailed in Section 2 and represents the main original result of this work.
By comparison, in Ref.~\cite{Feruglio:2016gvd} only the results strictly needed for the discussion of some physical processes
were presented, without discussing the issues related to the derivation, such as the RGE
flow, the matching conditions, or the consistency, such as the independence of the physical results
from the running scale. For completeness, in section 3, we discuss the phenomenological implications of our findings focusing on 
both Z-pole observables and low-energy observables, such as $\tau$ and $B$ meson decays, extending the concise discussion
made in Ref.~\cite{Feruglio:2016gvd}.
In section 4, we investigate the relevance of our results on specific classes of NP models such as minimal flavor violating models, 
$U(2)$ models and composite Higgs models. Our conclusions are presented in section 5.
\section{Theoretical framework}
\label{sec2}
If the new physics (NP) contributions originate at a scale much larger than the electroweak scale $v=246$ GeV, 
in the energy window above $v$ and below the NP mass scale, the NP effects can be
described by an effective Lagrangian invariant under the gauge group of the Standard Model (SM): 
\begin{align}
{\cal L}&={\cal L}_{SM}+{\cal L}_{NP}\vir\\
{\cal L}^{NP}&=\frac{1}{\Lambda^2}\sum_i C_i O_i+...\,,
\label{eq2:LNP}
\end{align}
where ${\cal L}_{SM}$ is the SM Lagrangian and $O_i$ are dimension six gauge invariant operators, $\Lambda$ represents 
the scale of NP and dots stand for higher dimension operators. 

In principle, 59 independent dimension six operators exist, which become 2499 when a completely general flavour structure 
is allowed \cite{Grzadkowski:2010es}. The discussion in full generality of the phenomenology arising from such a gigantic Lagrangian is clearly inconceivable and some additional assumptions are necessary. 
Here we assume that at a scale $\Lambda$, higher than the electroweak scale, the NP effects are fully described by the semileptonic operators 
$O^{(1)}_{\ell q}$ and $O^{(3)}_{\ell q}$ of table~\ref{tab:operators}. Moreover, we further assume that a basis exists where NP affects only the third fermion
generation\footnote{More generally, this assumption can be relaxed to say that NP mainly interact with only one fermion 
generation and interactions with the other two generations can be neglected.}. As a result, our effective Lagrangian at the 
scale $\Lambda$ is given by
\begin{equation}
{\cal L}_{NP}^0(\Lambda)=\frac{1}{\Lambda^2}
\left(
C_1~ \bar q'_{3L}\gamma^\mu q'_{3L}~ \bar \ell'_{3L}\gamma_\mu \ell'_{3L}+C_3~\bar q'_{3L}\gamma^\mu \tau^a q'_{3L}~ 
\bar \ell'_{3L}\gamma_\mu \tau^a \ell'_{3L}
\right)~,
\label{LNP}
\end{equation}
where primed fields are meant to be in a generic interaction basis. 
The above setup is the most natural one to accommodate simultaneously and in a correlated way charged- and neutral-current anomalies. 
Moreover, it is favoured by global fit analyses of $b\to s\ell^+\ell^-$ data~\cite{Hiller:2014yaa} 
including the very recent experimental result for $R^{\mu/e}_{K^*}$~\cite{Ciuchini:2017mik}.

%
At the electroweak scale $m_{EW}$, additional operators will arise from Lagrangian (\ref{LNP}), due to the well-known phenomenon of operator mixing. Here we will consider a reduced set of gauge-invariant operators involved in this RGE flow. Such set is summarised in Table \ref{tab:operators}, where one can recognize in $[O^{(1)}_{\ell q}]_{3333}$, $[O^{(3)}_{\ell q}]_{3333}$ the two initial operators of eq.~(\ref{LNP}).
\begin{table}[t]
\centering
\begin{tabular}{| r  @{ = } l |  r  @{ = } l  |}
	\hline
	\multicolumn{2}{| l |}{Semileptonic operators:} 		& \multicolumn{2}{ l |}{Leptonic operators:} \\
	\hline
			$[O^{(1)}_{\ell q}]_{prst}$
			&$(\bar \ell'_{pL}\gamma_\mu \ell'_{rL})~(\bar q'_{sL}\gamma^\mu q'_{tL})$
&
			$[O_{\ell \ell}]_{prst}$
			&$(\bar \ell'_{pL}\gamma_\mu \ell'_{rL})~(\bar \ell'_{sL}\gamma^\mu \ell'_{tL})$
\\
			$[O^{(3)}_{\ell q}]_{prst}$
			&$(\bar \ell'_{pL}\gamma_\mu\tau^a \ell'_{rL})~(\bar q'_{sL}\gamma^\mu\tau^a q'_{tL})$
&
			$[O_{\ell e}]_{prst}$
			&$(\bar \ell'_{pL}\gamma_\mu \ell'_{rL})~(\bar e'_{sR}\gamma^\mu e'_{tR})$
\\
			$[O_{\ell u}]_{prst}$
			&$(\bar \ell'_{pL}\gamma_\mu \ell'_{rL})~(\bar u'_{sR}\gamma^\mu u'_{tR})$
&
			\multicolumn{2}{ c |}{~}
\\
			$[O_{\ell d}]_{prst}$
			&$(\bar \ell'_{pL}\gamma_\mu \ell'_{rL})~(\bar d'_{sR}\gamma^\mu d'_{tR})$
&
			\multicolumn{2}{ c |}{~}
\\
			$[O_{q e}]_{prst}$
			&$(\bar q'_{pL}\gamma_\mu q'_{rL})~(\bar e'_{sR}\gamma^\mu e'_{tR})$
&
			\multicolumn{2}{ c |}{~}
\\
	\hline
	\multicolumn{2}{| l |}{Vector operators:} 		& \multicolumn{2}{ l |}{Hadronic operators:} \\
	\hline
			$[O^{(1)}_{H\ell}]_{pr}$
			&$(\varphi^\dagger i \overleftrightarrow{D_\mu} \varphi)~(\bar \ell'_{pL}\gamma_\mu \ell'_{rL})$
&
			$[O^{(1)}_{q q}]_{prst}$
			&$(\bar q'_{pL}\gamma_\mu q'_{rL})~(\bar q'_{sL}\gamma^\mu q'_{tL})$
\\
			$[O^{(3)}_{H\ell}]_{pr}$
			&$(\varphi^\dagger i \overleftrightarrow{D^a_\mu} \varphi)~(\bar \ell'_{pL}\gamma_\mu \tau^a \ell'_{rL})$
&
			$[O^{(3)}_{q q}]_{prst}$
			&$(\bar q'_{pL}\gamma_\mu\tau^a q'_{rL})~(\bar q'_{sL}\gamma^\mu\tau^a q'_{tL})$
\\
			$[O^{(1)}_{H q}]_{pr}$
			&$(\varphi^\dagger i \overleftrightarrow{D_\mu} \varphi)~(\bar q'_{pL}\gamma_\mu q'_{rL})$
&
			$[O^{(1)}_{q u}]_{prst}$
			&$(\bar q'_{pL}\gamma_\mu q'_{rL})~(\bar u'_{sR}\gamma^\mu u'_{tR})$
\\
			$[O^{(3)}_{H q}]_{pr}$
			&$(\varphi^\dagger i \overleftrightarrow{D^a_\mu} \varphi)~(\bar q'_{pL}\gamma_\mu \tau^a q'_{rL})$
&
			$[O^{(1)}_{q d}]_{prst}$
			&$(\bar q'_{pL}\gamma_\mu q'_{rL})~(\bar d'_{sR}\gamma^\mu d'_{tR})$
\\
\hline
\end{tabular}
\caption{Minimal set of gauge-invariant operators involved in the RGE flow considered in this paper. Fields are in the interaction basis to maintain explicit $SU(2)\times U(1)$ gauge invariance. Our notation and conventions are as in \cite{Grzadkowski:2010es}.
\label{tab:operators}}
\end{table}
Some work is needed in order to extract phenomenological predictions from Lagrangian (\ref{LNP}). This section is devoted to these computations. First of all, ${\cal L}_{NP}^0(\Lambda)$ should be run down to the EW scale $m_{EW}$ and a connection between the primed interaction basis introduced above and the fermion mass basis should be established (see section 2.1). Next, to describe several processes of interest, the $W$, $Z$ bosons and the $t$ quark should be integrated out (see section 2.2). Finally, for some processes a further RGE running (due to QED only) from $m_{EW}$ to an energy scale of order 1 GeV is needed (see section 2.3).

\subsection{Electroweak renormalization group flow}
In our framework NP effects are dominated by the Lagrangian ${\cal L}^{NP}$ of eq. (\ref{LNP}) at the scale $\Lambda$, which is assumed to be larger than the electroweak scale. 
In the particular application we have in mind $\Lambda$ is of order TeV. At energies smaller than $\Lambda$ the RGEs renormalise the coefficients $C_{1,3}$ and give rise to additional operators not initially included in ${\cal L}^{NP}$.

The anomalous dimension of these operators are known to one-loop accuracy~\cite{Jenkins:2013wua,Alonso:2013hga} and we can solve the related RGEs in a leading logarithmic approximation. By using ${\cal L}^{NP}$ of eq. (\ref{LNP}) as initial condition at the scale $\Lambda$ we obtain the effective Lagrangian ${\cal L}_{eff}$ at the scale $m_W\le\mu<\Lambda$:
\be
{\cal L}_{eff}={\cal L}_{SM}+{\cal L}_{NP}^0+\delta{\cal L}_{SL}+\delta{\cal L}_{V}+\delta{\cal L}_{L}+\delta{\cal L}_{H}~,
\label{leff}
\ee
\begin{align}
\delta{\cal L}_{SL}=\Bigg\{&\left.\left(g_1^2 C_1-9 g_2^2 C_3\right)~(\bar \ell'_{3L}\gamma_\mu \ell'_{3L})~(\bar q'_{3L}\gamma_\mu q'_{3L}) 
\right.\nonumber\\*
&\left.-\frac{2}{9} g_1^2 C_1~(\bar \ell'_{3L}\gamma_\mu \ell'_{3L})~(\bar q'_{sL}\gamma_\mu q'_{sL})-\frac{2}{3} g_1^2 C_1~(\bar \ell'_{sL}\gamma_\mu \ell'_{sL})~(\bar q'_{3L}\gamma_\mu q'_{3L})\right.\nonumber\\
&\left.-\frac{1}{2}  \left[(Y_u^\dagger Y_u)_{s3} \delta_{3t}+\delta_{s3}(Y_u^\dagger Y_u)_{3t}\right]       C_1~(\bar \ell'_{3L}\gamma_\mu \ell'_{3L})~(\bar q'_{sL}\gamma_\mu q'_{tL})\right.\nonumber\\
&\left. \left[-3 g_2^2 C_1+\left(6 g_2^2 +g_1^2 \right) C_3\right]~(\bar \ell'_{3L}\gamma_\mu\tau^a \ell'_{3L})~(\bar q'_{3L}\gamma_\mu \tau^a  q'_{3L}) \right.\nonumber\\
&\left.- 2 g_2^2 C_3~(\bar \ell'_{3L}\gamma_\mu\tau^a  \ell'_{3L})~(\bar q'_{sL}\gamma_\mu \tau^a  q'_{sL})
-\frac{2}{3} g_2^2 C_3~(\bar \ell'_{sL}\gamma_\mu\tau^a  \ell'_{sL})~(\bar q'_{3L}\gamma_\mu\tau^a  q'_{3L})
\right.\nonumber\\
&\left.-\frac{1}{2}   \left[(Y_u^\dagger Y_u)_{s3} \delta_{3t}+\delta_{s3}(Y_u^\dagger Y_u)_{3t}\right]       C_3~(\bar \ell'_{3L}\gamma_\mu\tau^a\ell'_{3L})~(\bar q'_{sL}\gamma_\mu\tau^a q'_{tL})\right.\nonumber\\
&\left.-\frac{8}{9} g_1^2 C_1~(\bar \ell'_{3L}\gamma_\mu \ell'_{3L})~(\bar u'_{sR}\gamma^\mu u'_{sR})+2 (Y_u)_{s3}(Y_u^\dagger)_{3t}~ C_1~(\bar \ell'_{3L}\gamma_\mu \ell'_{3L})~(\bar u'_{sR}\gamma^\mu u'_{tR})
\right.\nonumber\\*
&+\frac{4}{9} g_1^2 C_1~(\bar \ell'_{3L}\gamma_\mu \ell'_{3L})~(\bar d'_{sR}\gamma^\mu d'_{sR})-\frac{4}{3} g_1^2 C_1~(\bar q'_{3L}\gamma_\mu q'_{3L})~(\bar e'_{sR}\gamma^\mu e'_{sR})
\Bigg\}\frac{L}{16\pi^2\Lambda^2}~,
\\
\delta{\cal L}_{V}=\bigg[&\left(-\frac{2}{3} g_1^2 C_1-6 y_t^2  \lambda^u_{33} C_1\right)~ (\varphi^\dagger i \overleftrightarrow{D_\mu} \varphi)~(\bar \ell'_{3L}\gamma_\mu \ell'_{3L})\nonumber\\*
&+\left(-2 g_2^2 C_3+6 y_t^2  \lambda^u_{33}C_3\right)~ (\varphi^\dagger i \overleftrightarrow{D^a_\mu} \varphi)~(\bar \ell'_{3L}\gamma_\mu \tau^a \ell'_{3L})\nonumber\\*
&+\frac{2}{3} g_1^2 C_1~(\varphi^\dagger i \overleftrightarrow{D_\mu} \varphi)~(\bar q'_{3L}\gamma_\mu q'_{3L})-\frac{2}{3} g_2^2 C_3~(\varphi^\dagger i \overleftrightarrow{D^a_\mu} \varphi)~(\bar q'_{3L}\gamma_\mu \tau^a q'_{3L})
\bigg]\frac{L}{16\pi^2\Lambda^2}~,
\label{LNPewV}\\
\delta{\cal L}_{L}=\bigg[& \left(\frac{2}{3}g_1^2 C_1+2 g_2^2 C_3\right)~(\bar \ell'_{3L}\gamma_\mu \ell'_{3L})~(\bar \ell'_{sL}\gamma_\mu \ell'_{sL})
-4 g_2^2 C_3~ (\bar \ell'_{3L}\gamma_\mu \ell'_{sL})(\bar \ell'_{sL}\gamma_\mu \ell'_{3L})
			\nonumber\\*
&+\frac{4}{3} g_1^2 C_1~(\bar \ell'_{3L}\gamma_\mu \ell'_{3L})~(\bar e'_{sR}\gamma_\mu e_{sR})\bigg]\frac{L}{16\pi^2\Lambda^2}~,
\label{LNPewL}\\
\delta{\cal L}_{H}=\bigg[&\frac{2}{9} g_1^2 C_1~ (\bar q'_{3L}\gamma_\mu q'_{3L})~ (\bar q'_{sL}\gamma^\mu q'_{sL})-\frac{2}{3} g_2^2 C_3~ (\bar q'_{3L}\gamma_\mu\tau^a q'_{3L})~(\bar q'_{sL}\gamma^\mu\tau^a q'_{sL})\nonumber\\*
&+\frac{8}{9} g_1^2 C_1~(\bar q'_{3L}\gamma_\mu q'_{3L})~(\bar u'_{sR}\gamma^\mu u'_{sR})-\frac{4}{9} g_1^2 C_1~(\bar q'_{3L}\gamma_\mu q'_{3L})~(\bar d'_{sR}\gamma^\mu d'_{sR})\bigg]\frac{L}{16\pi^2\Lambda^2}~,
\end{align}
where
\be
L=\log\frac{\Lambda}{\mu}~,
\ee
and a sum over repeated indices is understood. In the above expressions we neglected all the Yukawa interactions but that of the top quark.

\subsubsection{Rotation to the mass basis}
\label{sec_massbasis}

In full generality, we can move to the mass basis by means of the unitary transformations
\begin{align}
\begin{aligned}
u'_L&=V_u u_L \\
\nu'_L&=U_e \nu_L
\end{aligned}
&&
\begin{aligned}
d'_L&=V_d d_L \\
e'_L&=U_e e_L
\end{aligned}
&&
\begin{aligned}
V_u^\dagger V_d=&V_{CKM}~, \\
~
\end{aligned}
\end{align}
where $V_{CKM}$ is the CKM mixing matrix and neutrino masses have been neglected.
We can now define the flavour matrices which parametrize the flavour structure of Lagrangian (\ref{leff}) in the mass basis:
%
%
\begin{align}
\begin{aligned}
\bar u'_{3}\gamma_\mu u'_{3}			&=	\lambda^u_{ij} \, \bar u_{i}\gamma_\mu u_{j}			\\
\bar d'_{3}\gamma_\mu d'_{3}			&= 		\lambda^d_{ij} \, \bar d_{i}\gamma_\mu d_{j}			\\
\bar u'_{3}\gamma_\mu d'_{3}			&= 		\lambda^{ud}_{ij} \, \bar u_{i}\gamma_\mu d_{j}	\\
\bar e'_{3}\gamma_\mu e'_{3}	&= 		\lambda^{e}_{ij} \, \bar e_{i}\gamma_\mu e_{j} \\
\bar \nu'_{3}\gamma_\mu \nu'_{3}	&= 	\lambda^{e}_{ij} \, \bar \nu_{i}\gamma_\mu \nu_{j}\\
\bar e'_{3}\gamma_\mu \nu'_{3}	&= 	\lambda^{e}_{ij} \, \bar e_{i}\gamma_\mu \nu_{j}
\end{aligned}
&&
\begin{aligned}
&\lambda^u_{ij}=V_{u3i}^*V_{u3j} 		&	\\
&\lambda^d_{ij}=V_{d3i}^*V_{d3j} 		&		\\
&\lambda^{ud}_{ij}=V_{u3i}^*V_{d3j} 	&		\\
& & \\
&\lambda^e_{ij}=U_{e3i}^*U_{e3j} ~.	&
& &
\end{aligned}
&&
\label{lambdas}
\end{align}
These matrices are redundant, since the following relations hold:
\begin{align}
&&
\lambda^u&=V_{CKM} \lambda^d V_{CKM}^\dagger
&
\lambda^{ud}&=V_{CKM}\lambda^d   ~.
&&
\end{align}
We also observe that $\lambda^{f}$ $(f=u,d,e)$ are hermitian matrices, satisfying $\lambda^f\lambda^f=\lambda^f$ and ${\rm tr}\lambda^f=1$ which shows they are projectors with one eigenvalue equal to one.  All the $\lambda$'s satisfy
\be
\sum\limits_{i,j=1}^3 |\lambda_{ij}|^2 =1~~~,
\ee
so that  all the elements $|\lambda_{ij}|$ should be smaller or equal to one. A useful and general parametrisation of $\lambda^f$ $(f=u,d,e)$ is:
\be
\lambda^f=
\frac{1}{1+|\alpha_f|^2+|\beta_f|^2}
\left(
\begin{array}{ccc}
|\alpha_f|^2&\alpha_f \bar\beta_f&\alpha_f\\
\bar\alpha_f \beta_f& |\beta_f|^2&\beta_f\\
\bar\alpha_f&\bar\beta_f&1
\end{array}
\right)~,
\label{parametrization}
\ee
where $\alpha_f$ and $\beta_f$ are complex numbers. Such general parametrisation directly follows from our assumption that a basis exists 
where NP affects only one fermion generation.
In summary the free parameters of our Lagrangian are the ratios $(C_{1,3})/\Lambda^2$ and the two matrices $\lambda^d$ and $\lambda^e$. 
From now on, we will work on the mass basis just defined. In the mass basis our starting point, eq. (\ref{LNP}), reads:
\begin{align}
{\cal L}_{NP}^0(\Lambda)=\frac{\lambda^e_{kl}}{\Lambda^2}\left[\right.&(C_1+C_3)~\lambda^u_{ij}~\bar u_{Li}\gamma^\mu u_{Lj}~ \bar\nu_{Lk}\gamma_\mu \nu_{Ll}+(C_1-C_3)~\lambda^u_{ij}~\bar u_{Li} \gamma^\mu u_{Lj}~\bar e_{Lk}\gamma_\mu e_{Ll}~+\label{lag1}\nn\\
&(C_1-C_3)~\lambda^d_{ij}~\bar d_{Li} \gamma^\mu d_{Lj}~ \bar\nu_{Lk}\gamma_\mu \nu_{Ll}+(C_1+C_3)~\lambda^d_{ij}~\bar d_{Li} \gamma^\mu d_{Lj}~\bar e_{Lk}\gamma_\mu e_{Ll}~+\\
&2 C_3~\left(\lambda^{ud}_{ij}~\bar u_{Li} \gamma^\mu d_{Lj}~\bar e_{Lk}\gamma_\mu \nu_{Ll}+h.c.\right)\left.\right]~.\nn
\end{align}
%

\subsubsection{Modified Z and W couplings}
Once performed the RGE running of Lagrangian (\ref{LNP}) down to $m_{EW}$ and discussed the rotation to the mass basis, we are already able to examine some phenomenology arising from Lagrangian (\ref{LNP}) and in particular from $\delta \cL_V$.
Indeed, one of its effects is the modification of the couplings of gauge vector bosons $W$ and $Z$ to fermions. 

The interaction Lagrangian between $Z,W$ and fermions can be expressed as:
\begin{align}
{\cal L}_{W,Z}=-\frac{g_2}{c_\theta}~ Z_\mu \, J_Z^\mu
-\frac{g_2}{\sqrt{2}}\bigg( W^+_\mu J^{-\mu}+\hc\bigg)  ~,
\label{lwz}
\end{align}
where $c_\theta = \cos\theta_W$ and
\begin{align}
J_Z^\mu &=\sum_f \left[g_{fL}^{ij} \bar f_{Li}\gamma^\mu f_{Lj}+g_{fR}^{ij}\bar f_{Ri}\gamma^\mu f_{Rj}\right] ~, \\
J^{-\mu} &=g_{\ell}^{ij}~\bar \nu_{Li}\gamma^\mu e_{Lj}+g_{q}^{ij}~\bar u_{Li}\gamma^\mu d_{Lj} ~.
\end{align}
The effective coupling constants include both the SM and the NP contributions:
\begin{align}
&&
g_{fL,R}&=g^{SM}_{fL,R}+\Delta g_{fL,R}
&
g_{\ell,q}&=g^{SM}_{\ell,q}+\Delta g_{\ell,q}~,
&&
\label{effWZc}
\end{align}
where the SM part is given by:
\begin{align}
(g_{fL}^{SM})^{ij} &= \left(T^f_{3L}-Q^f  s^2_\theta \right)\, \delta_{ij}
&
(g_{fR}^{SM})^{ij} &= -Q^f s^2_\theta\, \delta_{ij} ~,
\end{align}
\begin{align}
(g_\ell^{SM})^{ij}&=\delta_{ij}
&
(g_q^{SM})^{ij} &=(V_{CKM})_{ij}~,
\end{align}
where $s_\theta = \sin\theta_W$.
For generic Wilson coefficients of the vector operators $[O^{(1,3)}_{Hl}]_{pr}$ and $[O^{(1,3)}_{Hq}]_{pr}$,  the NP contributions read
\begin{equation}
\begin{aligned}
\Delta g_{\nu L}^{ij}&=-\frac{v^2}{2\Lambda^2}\left[ U_e^+\left(C^{(1)}_{Hl}-C^{(3)}_{Hl}\right) U_e\right]_{ij} \\
\Delta g_{e L}^{ij}&=-\frac{v^2}{2\Lambda^2}\left[ U_e^+\left(C^{(1)}_{Hl}+C^{(3)}_{Hl}\right) U_e\right]_{ij}  \\
\Delta g_{u L}^{ij}&=-\frac{v^2}{2\Lambda^2}\left[ V_u^+\left(C^{(1)}_{Hq}-C^{(3)}_{Hq}\right) V_u\right]_{ij}  \\
\Delta g_{d L}^{ij}&=-\frac{v^2}{2\Lambda^2}\left[ V_d^+\left(C^{(1)}_{Hq}+C^{(3)}_{Hq}\right) V_d\right]_{ij}   \\
\Delta g_{f R}^{ij}&=0~~~~~(f=\nu,e,u,d)  \\
\Delta g_{\ell}^{ij}&=\frac{v^2}{\Lambda^2} \left[ U_e^+ C^{(3)}_{Hl} U_e\right]_{ij}    \\
\Delta g_q^{ij}&=\frac{v^2}{\Lambda^2} \left[ V_u^+ C^{(3)}_{Hq} V_d\right]_{ij}  ~~~.
\end{aligned}
\label{deltagen}
\end{equation}
For our case they can be directly derived from Lagrangian (\ref{LNPewV}):
\begin{equation}
\begin{aligned}
\Delta g_{\nu L}^{ij}&=\frac{v^2}{\Lambda^2}\frac{L}{16\pi^2}\big(\frac{1}{3} g_1^2 C_1-g_2^2 C_3+3y_t^2 \lambda^u_{33} (C_1+C_3)\big)\lambda^e_{ij}  \\
\Delta g_{e L}^{ij}&=\frac{v^2}{\Lambda^2}\frac{L}{16\pi^2}\big(\frac{1}{3} g_1^2 C_1+g_2^2 C_3+3y_t^2 \lambda^u_{33} (C_1-C_3)\big)\lambda^e_{ij}  \\
\Delta g_{u L}^{ij}&=-\frac{v^2}{\Lambda^2}\frac{L}{16\pi^2}\frac{1}{3}\big( g_1^2 C_1+g_2^2 C_3\big)\lambda^u_{ij}  \\
\Delta g_{d L}^{ij}&=-\frac{v^2}{\Lambda^2}\frac{L}{16\pi^2}\frac{1}{3}\big( g_1^2 C_1-g_2^2 C_3\big)\lambda^d_{ij}\\
\Delta g_{f R}^{ij}&=0~~~~~(f=\nu,e,u,d)  \\
\Delta g_{\ell}^{ij}&=\frac{v^2}{\Lambda^2}\frac{L}{16\pi^2}(-2 g_2^2 C_3+6 y_t^2 \lambda^u_{33} C_3)\lambda^{e}_{ij}  \\
\Delta g_q^{ij}&=-\frac{v^2}{\Lambda^2}\frac{L}{16\pi^2}\frac{2}{3} g_2^2 C_3\lambda^{ud}_{ij}~~~.
\end{aligned}
\label{deltas}
\end{equation}

\begin{figure}[!ht]
\centering
\includegraphics[width=0.25\textwidth]{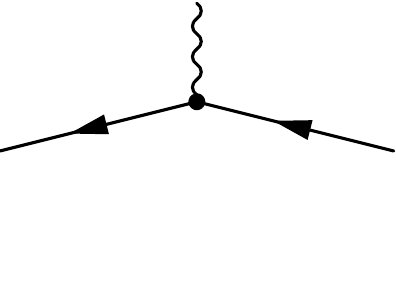}
\hspace{1.5 truecm}
\includegraphics[width=0.25\textwidth]{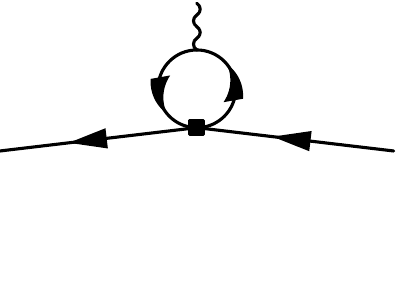}
\vspace{-1.0 truecm}
\caption{Diagrams contributing to the decay of the $Z$ into a lepton pair. On the left, the contribution from ${\cal L}_{W,Z}$, eq. (\ref{lwz}). On the right the one-loop contribution originating from the four-fermion interactions 
contained in ${\cal L}^0_{NP}(\Lambda)$, eq. (\ref{LNP}), denoted by a square. The dependence on the renormalization scale $\mu$ cancels in the sum.}
\label{fig1}
\end{figure}

These effective couplings depends on the renormalisation scale $\mu$. Such a dependence cancels when we compute the amplitudes for the relevant processes. For example, to compute the decay amplitude of the $Z$ boson into a lepton pair, the contribution from the effective Lagrangian ${\cal L}_{W,Z}$, which is formally a tree-level term, should be combined with that 
of the one-loop diagram arising from the four-fermion interactions contained in ${\cal L}^{NP}$, with the $Z$ boson on the 
mass-shell attached to the quark legs, as shown in fig. \ref{fig1}.

This has the effect of replacing the couplings of eq. (\ref{effWZc}) with
\be
g_{fL,R}=g^{SM}_{fL,R}+\delta g_{fL,R}~~~~~~~~~~~~~(f=\nu,e)~~~,
\ee
where
\begin{align}
\label{newgnuL}
\delta g_{\nu L}^{ij}&=\frac{3}{8 \pi^2}
\left\{\frac{m_t^2}{\Lambda^2} \lambda^u_{33} (C_1+C_3) \left(\log\frac{\Lambda}{m_Z} - \frac{I_1}{2}\right)\right.
+\frac{m_Z^2}{\Lambda^2} \lambda^u_{33} (C_1+C_3)(1-\frac{4}{3}s^2_\theta)(I_2-I_3)\nn\\
&+\left.\frac{m_Z^2}{\Lambda^2}\left[(C_1+C_3)(1-\frac{4}{3}s^2_\theta)+(C_1-C_3)(-1+\frac{2}{3}s^2_\theta)\right]
\left(I_3 - \frac{1}{3}\log\frac{\Lambda}{m_Z}\right)
\right\}\lambda^e_{ij}~,
\end{align}
and
\begin{align}
\label{newgeL}
\delta g_{eL}^{ij}&=\frac{3}{8 \pi^2}
\left\{\frac{m_t^2}{\Lambda^2} \lambda^u_{33} (C_1-C_3) \left(\log\frac{\Lambda}{m_Z} - \frac{I_1}{2}\right)\right.
+\frac{m_Z^2}{\Lambda^2} \lambda^u_{33} (C_1-C_3)(1-\frac{4}{3}s^2_\theta)(I_2-I_3)\nn\\
&+\left.\frac{m_Z^2}{\Lambda^2}\left[(C_1-C_3)(1-\frac{4}{3}s^2_\theta)+(C_1+C_3)(-1+\frac{2}{3}s^2_\theta)\right]
\left(I_3 - \frac{1}{3}\log\frac{\Lambda}{m_Z}\right)
\right\}\lambda^e_{ij}~,
\end{align}
where $I_i$ $(i=1,2,3)$ are finite and renormalization scale independent quantities defined as
\begin{align}
I_1 &=  \int_0^1 dx \log\frac{m_t^2-m_Z^2~x(1-x)}{m_Z^2}~,\\
I_2 &= \int_0^1 dx~x(1-x) \log\frac{m_t^2- m_Z^2~ x (1-x)}{m_Z^2}~,\\
I_3 &= \int_0^1 dx~ x(1-x)\log[x (x-1)]~.
\end{align}
%
Starting from the above expressions, we find the following approximate results
\begin{align}
\delta g_{\nu L}^{ij}
&\approx\frac{10^{-3}}{\Lambda^2}\left\{(2.1\!+\!1.1\log\Lambda)\lambda^u_{33}(C_1\!+\!C_3)-0.52 C_3
-i [0.11\lambda^u_{33} (C_1\!+\!C_3)-0.25 C_3]\right\}\lambda^e_{ij}~,
\nn\\
\delta g_{eL}^{ij}&\approx
\frac{10^{-3}}{\Lambda^2}\left\{(2.1\!+\!1.1\log\Lambda)\lambda^u_{33}(C_1\!-\!C_3)+0.52 C_3
-i[0.11\lambda^u_{33} (C_1\!-\!C_3)+0.25 C_3]\right\}\lambda^e_{ij}\nn~,
\end{align}
where $\Lambda$ should be evaluated in TeV.

The scale dependence of the RGE contribution cancels with that of the matrix element dominated by a quark loop. We kept non-vanishing only the top mass, 
neglecting corrections of order $m_q^2/(16\pi^2\Lambda^2)$ when $q=u,d,c,s,b$. These quarks are responsible for the imaginary part of the effective coupling constants.
\vskip 1.cm
\noindent

\subsection{Effective theory below \texorpdfstring{$t$, $W$ and $Z$}{t, W, Z} thresholds}

At the electroweak scale we should integrate out the heavy particles $W$, $Z$ and the top quark $t$. We will not distinguish 
among the mass scales $m_W$, $m_Z$ and $m_t$, which we will approximately identify with a common scale $m_{EW}$.
Below such scale, we should use a new effective Lagrangian ${\cal L}_{eff}^{EW}$, made of four fermion operators whose 
Wilson coefficients are determined by a matching procedure, i.e.~by requiring that amplitudes computed with ${\cal L}_{eff}$, 
eq. (\ref{leff}), and ${\cal L}_{eff}^{EW}$ are the same at the scale $m_{EW}$. We work in leading log approximation (LLA) 
with one-loop accuracy. Therefore, matching conditions should include both contributions from tree-level topologies with 
RGE-improved dimension six operator insertions and contributions from one-loop topologies with tree-level-accurate dimension six 
operator insertions.

One of the steps required by this procedure is integrating out the massive gauge bosons $W$, $Z$. In particular, from the 
tree-level exchange of $W$ and $Z$ we get:
\begin{align}
\delta \cL_{V}^{EW} 
										= -\frac{2}{v^2} \left(  J_Z^{SM\mu} J^{SM}_{Z\mu}+J^{SM+\mu}J^{SM-}_\mu
																						+ 2 J_{Z \,\mu}^{SM} \Delta J_{Z}^\mu
																						+\left( J^{SM+}_\mu \Delta J^{-\,\mu} +\hc \right)
																						   \right) ~,
\label{LNPEW}
\end{align}
where:
\begin{align}
J_{Z\,\mu}^{SM} &= \sum_f \left[(g_{fL}^{SM})^{ij}\bar f_{Li}\gamma_\mu f_{Lj}
											+ (g_{fR}^{SM})^{ij}\bar f_{Ri}\gamma_\mu f_{Rj}\right]\,,
&
\Delta J_Z^\mu &= \sum_f \Delta g_{f L}^{ij} \bar f_{Li}\gamma^\mu f_{Lj}	\,,
\\
J^{SM-}_\mu &= (g_\ell^{SM})^{ij} \, \bar \nu_{Li}\gamma_\mu e_{Lj}+(g_q^{SM})^{ij} \, \bar u_{Li}\gamma_\mu d_{Lj}\,,
&
\Delta J^{-}_\mu &= \Delta g_{\ell}^{ij} \, \bar \nu_{Li}\gamma_\mu e_{Lj}+\Delta g_q^{ij} \,\bar u_{Li}\gamma_\mu d_{Lj}\,,
\end{align}
where the contributions proportional to $\Delta J_Z^\mu$, $\Delta J^{\pm}_\mu$ in eq. (\ref{LNPEW}) are of order one-loop.

\begin{figure}[t]
\centering
\begin{minipage}{0.15\textwidth}
\includegraphics[width=\textwidth]{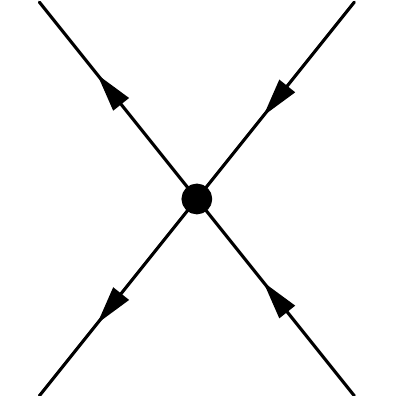}
\caption*{(2.a)}
\end{minipage}
\begin{minipage}{0.15\textwidth}
\includegraphics[width=\textwidth]{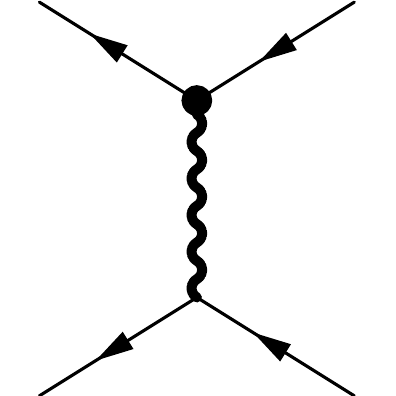}
\caption*{(2.b)}
\end{minipage}
\begin{minipage}{0.15\textwidth}
\includegraphics[width=\textwidth]{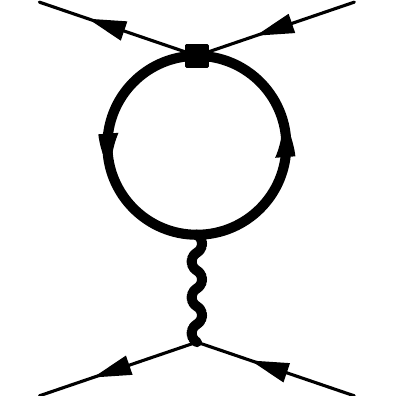}
\caption*{(2.c)}
\end{minipage}
\begin{minipage}{0.15\textwidth}
\includegraphics[width=\textwidth]{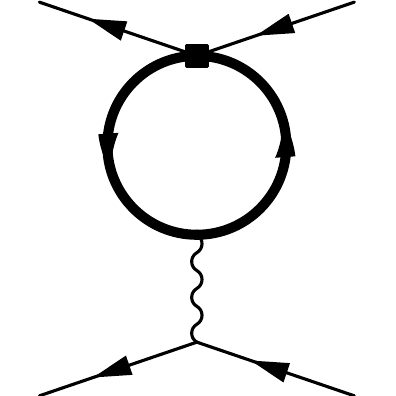}
\caption*{(2.d)}
\end{minipage}
\begin{minipage}{0.15\textwidth}
\includegraphics[width=\textwidth]{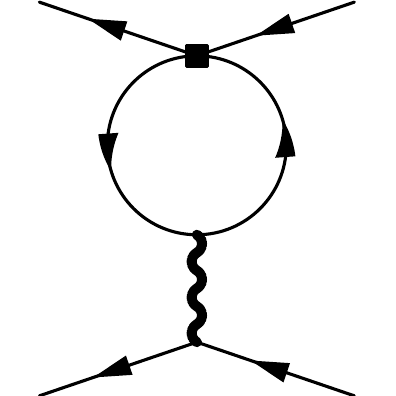}
\caption*{(2.e)}
\end{minipage}
\begin{minipage}{0.15\textwidth}
\includegraphics[width=\textwidth]{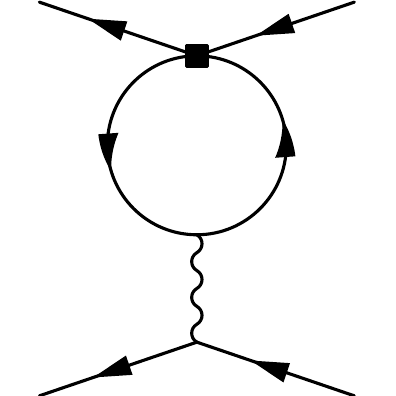}
\caption*{(2.f)}
\end{minipage}
\caption{Diagrams contributing to the Wilson coefficient of $(\bar e_{iL} \gamma_\mu e_{jL}) \, (\bar e_{kL}\gamma^\mu e_{nL})$ above the electroweak scale $m_{EW}$, computed with Lagrangian ${\cal L}_{eff}$. Thick (thin) lines denote
heavy (light) fields. 
Diagram (2.a) comes from $\delta {\cal L}_L$, (2.b) from $\delta {\cal L}_V$. The diagrams (2.c)--(2.f) come from the four-fermion interactions 
of  ${\cal L}^0_{NP}(\Lambda)$, eq. (\ref{LNP}), denoted by a square. They cancel the dependence on the renormalization scale $\mu$ of the diagrams (2.a), (2.b).}
\label{above}
\end{figure}

\begin{figure}[t]
\centering
\begin{minipage}{0.15\textwidth}
\includegraphics[width=\textwidth]{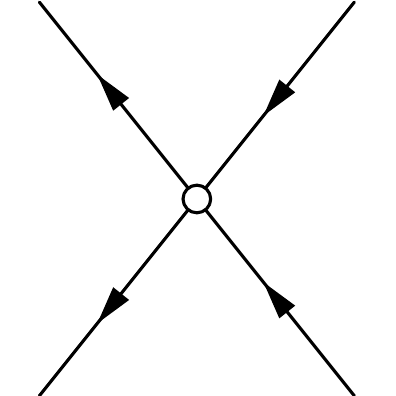}
\caption*{(3.a)}
\end{minipage}
~~~
\begin{minipage}{0.15\textwidth}
\includegraphics[width=\textwidth]{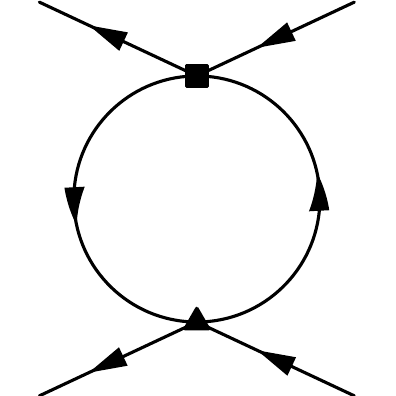}
\caption*{(3.b)}
\end{minipage}
~~~
\begin{minipage}{0.15\textwidth}
\includegraphics[width=\textwidth]{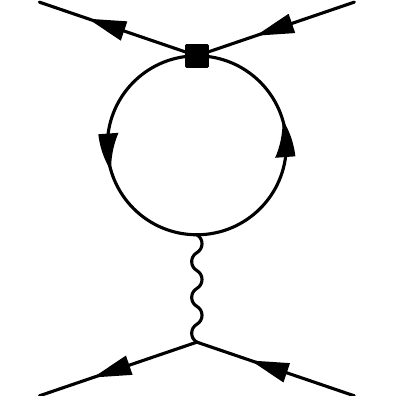}
\caption*{(3.c)}
\end{minipage}
\caption{Diagrams contributing to the Wilson coefficient of $(\bar e_{iL} \gamma_\mu e_{jL}) \, (\bar e_{kL}\gamma^\mu e_{nL})$ below the electroweak scale $m_{EW}$. Only light fields (thin lines) are present. 
The four-fermion interaction (empty circle) of diagram (3.a) is the unknown of the matching procedure. The triangle in diagram (3.b) denotes the SM Fermi interaction.
In the matching condition, the diagram (3.b) is canceled by the diagram (2.e) and the diagram (3.c) is canceled by the diagram (2.f).}
\label{below}
\end{figure}

As an explicit example of matching procedure, we discuss in detail the case of the four-fermion operator $(\bar e_{iL} \gamma_\mu e_{jL}) \, (\bar e_{kL}\gamma^\mu e_{nL})$. Above the electroweak scale $m_{EW}$, the amplitude of the processes controlled by this operator is given by the 
sum of the diagrams in fig.~\ref{above}. Diagrams (2.a) and (2.b) come from $\delta {\cal L}_L$ (eq.~(\ref{LNPewL})) and $\delta \cL_{V}^{EW}$ (eq.~(\ref{LNPEW})), respectively. They depend on the renormalization scale $\mu$ through the Wilson coefficients of the interactions displayed 
as a full circle. This dependence is cancelled by the diagrams (2.c)--(2.f). 
Below the electroweak scale, $W$, $Z$ and $t$ have been removed from the effective Lagrangian and the amplitude is given by the sum of the diagrams in fig. \ref{below}. The diagram (3.a) represents our unknown. By equating the amplitudes above and below $m_{EW}$, we find that (3.a) 
is given by the sum of (2.a)--(2.d), since (3.b) and (3.c) are canceled by (2.e) and (2.f) respectively:
\begin{equation}
{\cal A}_{3a}={\cal A}_{2a}+{\cal A}_{2b}+{\cal A}_{2c}+{\cal A}_{2d}  ~,
\label{mc}
\end{equation}
where we have exploited ${\cal A}_{3c}={\cal A}_{2f}$ and, at first order in $G_F$, ${\cal A}_{3b}={\cal A}_{2e}$. From eqs.~(\ref{LNPewL}), (\ref{lambdas}), (\ref{deltas}) and (\ref{LNPEW}) we get:
\begin{align}
{\cal A}_{2a} &= \frac{2}{3}\left( g_1^2 C_1-3 g_2^2 C_3
\right)\frac{1}{16\pi^2\Lambda^2}\log\frac{\Lambda}{\mu}	
					\cdot \lambda^e_{ij} \delta_{kn} ~,\\
{\cal A}_{2b} &= (1-2\sin^2\theta_W)\left[\frac{2}{3}\left(g_1^2 C_1+3 g_2^2 C_3\right)+6y_t^2 \lambda^u_{33}(C_1-C_3)\right]\frac{1}{16\pi^2\Lambda^2}\log\frac{\Lambda}{\mu} 
					\cdot \lambda^e_{ij} \delta_{kn} ~.
\end{align}
An explicit computation of diagrams (2.c) and (2.d) at the leading logarithmic order gives:
\begin{align}
{\cal A}_{2c}&=6(1-2\sin^2\theta_W) y_t^2 \lambda^u_{33}(C_1-C_3)\frac{1}{16\pi^2\Lambda^2}\log\frac{\mu}{m_t}   \cdot \lambda^e_{ij} \delta_{kn} +... ~,\\
{\cal A}_{2d}&=\frac{8}{3} e^2 \lambda^u_{33} (C_1-C_3)\frac{1}{16\pi^2\Lambda^2}\log\frac{\mu}{m_t}			\cdot \lambda^e_{ij} \delta_{kn} +... ~,
\end{align}
where dots stand for $\mu$-independent finite contributions. Hereafter, we will not include such contributions in our analysis, 
as finite terms of the same size could originate from UV completions of our setup. Therefore our quantitative conclusions rely on 
the assumption that the logarithmic RGE-induced terms dominate over the finite ones. Given the relatively short range where 
the running takes place, it is not guaranteed that such dominance holds and the possibility of cancellations in physical observables 
cannot be excluded.
Putting all together, we get
\begin{align}
{\cal A}_{3a}=
\frac{1}{16\pi^2\Lambda^2}
&\left[
 				-12(-\frac12+\sin^2\theta_W)y_t^2 \lambda^u_{33}(C_1-C_3) \log\frac{\Lambda}{m_t} 
 				+ \frac{4}{3} e^2(C_1-3C_3)\log\frac{\Lambda}{m_t} \right. \nn\\
 &\left. +\frac{4}{3} e^2(C_1-C_3) \lambda^u_{33} \log\frac{\mu}{m_t}\right] \cdot \lambda^e_{ij} \delta_{kn}  ~.
 \label{eq:sr1}
\end{align}
At a scale $\mu$ just below $m_{EW}\approx m_t$, we can drop the last term of eq.~(\ref{eq:sr1}). In this way we end up with our final result for the Wilson coefficient of $(\bar e_{iL} \gamma_\mu e_{jL}) \, (\bar e_{kL}\gamma^\mu e_{nL})$ below $m_{EW}$, reported in table \ref{tab:4fL}.

To derive the full ${\cal L}_{eff}^{EW}$, we should repeat this procedure for each four-fermion operator arising from $\cL_{eff}$.  In practice, as we have seen in the previous example, this amounts to:
\begin{enumerate}
\item move to the fermion mass basis in $({\cal L}_{eff}-\delta{\cal L}_{V})={\cal L}_{SM}+{\cal L}_{NP}^0+\delta{\cal L}_{SL}+\delta{\cal L}_{L}+\delta{\cal L}_{H}$, removing every operator featuring a $W$, $Z$ or $t$ fields;
\item  add the result to the term originating from tree-level $W$ and $Z$ exchange, $\delta{\cal L}_{V}^{EW}$, eq.~(\ref{LNPEW});
\item  set the scale $\mu$ to $m_{EW}\approx m_{W,Z,t}$, in this way accounting for the one-loop topologies contributions.
\end{enumerate}

Our final result reads:
\begin{align}
\cL_{eff}^{EW} = \cL_{SM}' + \cL_{NP}^0
										+\frac{1}{16\pi^2\Lambda^2}\log\frac{\Lambda}{m_{EW}}~\sum_i \xi_i Q_i ~,
\label{LEWlow}
\end{align}
where $\cL_{SM}'$ is the SM Lagrangian where $W$ and $Z$ has been integrated out (Fermi theory), and the systematic omission on the right-hand side of the $t_{L,R}$ quark fields is understood.
The four-fermion operators $Q_i$ and their Wilson coefficients $\xi_i$ are given in Table \ref{tab:4fL}--\ref{ta:4fCC} for the semileptonic and purely leptonic type.

\begin{table}[tp] 		
\centering
\begin{tabular}{| l | l |}
		\hline
		\multicolumn{1}{| c |}{$Q_i$} & \multicolumn{1}{c |}{$\xi_i$}\\
		\hline
		&\\[-19pt]
		\hline
		$(\bar\nu_{i L}\gamma_\mu \nu_{j L})~(\bar \nu_{k L}\gamma^\mu  \nu_{n L})$
					&
					$ \phantom{+} \lambda^e_{ij} \, \delta_{kn}    
								\left[  -6 y_t^2 \lambda^u_{33}(C_1+C_3)  \right]$ \\ [3pt]
		\hline
		$(\bar\nu_{i L}\gamma_\mu \nu_{j L})~(\bar e_{k L}\gamma^\mu  e_{n L})$
					&
					$ \phantom{+} \lambda^e_{ij} \, \delta_{kn}
								\left[  \frac43e^2(C_1+3C_3)  -12\, (-\frac12+s^2_\theta) \, y_t^2 \lambda^u_{33}(C_1+C_3)  \right]$ \\
					&
					$ + \delta_{ij} \, \lambda^e_{kn}
								\left[  -6 y_t^2 \lambda^u_{33}(C_1-C_3)  \right]$ \\[3pt]
		\hline
		$(\bar\nu_{i L}\gamma_\mu \nu_{j L})~(\bar e_{k R}\gamma^\mu  e_{n R})$
					&
					$ \phantom{+} \lambda^e_{ij} \, \delta_{kn}
								\left[  \frac43 e^2(C_1+3C_3)-12 \, s^2_\theta \, y_t^2 \lambda^u_{33}(C_1+C_3)  \right]$ \\ [3pt]
\hline
		$(\bar e_{i L}\gamma_\mu e_{j L})~(\bar e_{k L}\gamma^\mu  e_{n L})$
					&
					$ \phantom{+} \lambda^e_{ij} \, \delta_{kn}
								\left[  \frac43 e^2(C_1-3C_3)-12(-\frac12+s^2_\theta) \, y_t^2 \lambda^u_{33}(C_1-C_3)  \right]$ \\ [3pt]
		\hline
		$(\bar e_{i L}\gamma_\mu e_{j L})~(\bar e_{k R}\gamma^\mu  e_{n R})$
					&
					$ \phantom{+} \lambda^e_{ij} \, \delta_{kn}
								\left[  \frac43 e^2(C_1-3C_3)-12 \, s^2_\theta \, y_t^2 \lambda^u_{33}(C_1-C_3)  \right]$ \\ [3pt]
		\hline
		&\\[-19pt]
		\hline
		$(\bar\nu_{i L}\gamma_\mu e_{j L})~(\bar e_{k L}\gamma^\mu  \nu_{n L})$
					&
					$ (\lambda^e_{ij} \, \delta_{kn}+\delta_{ij} \,\lambda^e_{kn})
								\left[  -12\, y_t^2 \lambda^u_{33} C_3   \right]$ \\ [3pt]		
						\hline
\end{tabular}
\caption{Operators $Q_i$ and coefficients $\xi_i$ for the purely leptonic part of the effective Lagrangian ${\cal L}_{eff}^{EW}$. We set $\sin^2\theta_W\equiv s^2_\theta$.\label{tab:4fL}}
\end{table}

\begin{table}[tp] 		
\centering
\begin{tabular}{| c | l |}
		\hline
		$Q_i$& \multicolumn{1}{c |}{$\xi_i$}\\
		\hline
		&\\[-19pt]
		\hline
		$(\bar\nu_{i L}\gamma_\mu \nu_{j L}) \, (\bar u_{k L}\gamma^\mu  u_{n L})$
					&
					$ \phantom{+} \lambda^e_{ij} \, \lambda^u_{k n}    
								\left[  (g_1^2-3g_2^2)(C_1+C_3)  \right]$ \\
					&
					$ + \lambda^e_{ij} \, \delta_{kn}
								\left[  -\frac89 e^2(C_1+3C_3)-12(\frac12 - \frac23 s^2_\theta)~ y_t^2 \lambda^u_{33}(C_1+C_3)  \right]$ \\					
					&
					$ + \lambda^e_{ij} \, (\lambda^u_{k3}\delta_{3n} + \delta_{k3}\lambda^u_{3n})
								\left[ - \frac{1}{2}y_t^2(C_1+C_3)  \right]$ \\ [3pt]
		\hline
		$(\bar\nu_{i L}\gamma_\mu \nu_{j L}) \, (\bar u_{k R}\gamma^\mu  u_{n R})$
					&
					$ \phantom{+} \lambda^e_{ij} \, \delta_{kn} 
					 			\left[   -\frac89 e^2(C_1+3C_3)+8 s^2_\theta \, y_t^2 \lambda^u_{33}(C_1+C_3)  \right]  $ \\
					&
					$ + \lambda^e_{ij} \, \delta_{k3} \delta_{3n}    
								\left[  2 y_t^2 \lambda^u_{33}  C_1  \right]$ \\ [3pt]
		\hline
		$(\bar\nu_{i L}\gamma_\mu \nu_{j L}) \, (\bar d_{k L}\gamma^\mu  d_{n L})$
					&
					$ \phantom{+} \lambda^e_{ij} \, \lambda^d_{k n}    
								\left[  (g_1^2+3g_2^2)C_1-(g_1^2+15g_2^2)C_3  \right]$ \\
					&
					$ + \lambda^e_{ij} \, \delta_{kn}
								\left[  \frac49 e^2(C_1+3C_3)-12(-\frac12+\frac13 s^2_\theta) \, y_t^2 \lambda^u_{33}(C_1+C_3)  \right]$ \\					
					&
					$ + \lambda^e_{ij} \, ((\lambda^{ud \, \dagger})_{k3}V^{\mysmall{CKM}}_{3n} 
																		+ (V^{\mysmall{CKM}})^\dagger_{k3} \lambda^{ud}_{3n})
								\left[  -\frac{1}{2}y_t^2(C_1-C_3)  \right]$ \\ [3pt]
		\hline
		$(\bar\nu_{i L}\gamma_\mu \nu_{j L}) \, (\bar d_{k R}\gamma^\mu  d_{n R})$
					&
					$ \phantom{+} \lambda^e_{ij} \, \delta_{kn}
								\left[  \frac49 e^2(C_1+3C_3)-4 s^2_\theta \, y_t^2 \lambda^u_{33}(C_1+C_3)  \right]$ \\ [3pt]
		\hline	
\end{tabular}
\caption{Operators $Q_i$ and coefficients $\xi_i$ for the semileptonic part of the effective Lagrangian ${\cal L}_{eff}^{EW}$ involving neutrinos and neutral currents.
Generation indices run from 1 to 3, exception made for up-type quarks where $k,n = 1,2$. We set $\sin^2\theta_W\equiv s^2_\theta$.}
\end{table}

\begin{table}[tp] 		
\centering
\begin{tabular}{| c | l |}
		\hline
		$Q_i$& \multicolumn{1}{c |}{$\xi_i$}\\
		\hline
		&\\[-19pt]
		\hline
		$(\bar e_{i L}\gamma_\mu e_{j L}) \, (\bar u_{k L}\gamma^\mu  u_{n L})$
					&
					$    \phantom{+} \lambda^e_{ij} \, \lambda^u_{kn}
								\left[  (g_1^2+3g_2^2)C_1-(g_1^2+15g_2^2)C_3  \right]$ \\ 
					&
					$    + \lambda^e_{ij} \, \delta_{kn}
								\left[  -\frac89 e^2(C_1-3C_3)-12(\frac12-\frac32 s^2_\theta) y_t^2 \lambda^u_{33}(C_1-C_3)  \right]$ \\ 
					&
					$    + \delta_{ij} \, \lambda^u_{kn}
								\left[  -\frac{4}{3} e^2(C_1-C_3)  \right]$ \\ 
					&
					$    + \lambda^e_{ij} \, (\lambda^u_{k3}\delta_{3n} + \delta_{k3}\lambda^u_{3n})
								\left[  -\frac12 y_t^2(C_1-C_3)  \right]$ \\ [3pt]
		\hline
		$(\bar e_{i L}\gamma_\mu e_{j L}) \, (\bar u_{k R}\gamma^\mu  u_{n R})$
					&
					$    \phantom{+} \lambda^e_{ij} \, \delta_{kn}
								\left[   -\frac89 e^2(C_1-3C_3)+8 \, s^2_\theta \, y_t^2 \lambda^u_{33}(C_1-C_3)  \right]$ \\
					&
					$    + \lambda^e_{ij} \, \delta_{k3} \delta_{3n}
								\left[  2 y_t^2 \lambda^u_{33} C_1  \right]$ \\ 
 [3pt]
		\hline
		$(\bar e_{i R}\gamma_\mu e_{j R}) \, (\bar u_{k L}\gamma^\mu  u_{n L})$
					&
					$    \phantom{+} \delta_{ij} \, \lambda^u_{kn}
								\left[  -\frac43 e^2(C_1-C_3)  \right]$ \\ [3pt]
		\hline
		$(\bar e_{i L}\gamma_\mu e_{j L}) \, (\bar d_{k L}\gamma^\mu  d_{n L})$
					&
					$    \phantom{+} \lambda^e_{ij} \, \lambda^d_{kn}
								\left[  (g_1^2-3g_2^2)(C_1+C_3)  \right]$ \\ 
					&
					$    + \lambda^e_{ij} \, \delta_{kn}
								\left[  \frac49 e^2(C_1-3C_3)-12(-\frac12+\frac13 s^2_\theta)\, y_t^2 \lambda^u_{33}(C_1-C_3)  \right]$ \\ 
					&
					$    + \delta_{ij} \, \lambda^d_{kn}
								\left[  -\frac43 e^2(C_1+C_3)  \right]$ \\ 
					&
					$    + \lambda^e_{ij} \, ((\lambda^{ud \, \dagger})_{k3}V^{\mysmall{CKM}}_{3n} 
																		+ {({V^{\mysmall{CKM}}}^\dagger)_{k3}} \lambda^{ud}_{3n})
								\left[  -\frac12 \, y_t^2(C_1+C_3)  \right]$ \\ [3pt]
		\hline
		$(\bar e_{i L}\gamma_\mu e_{j L}) \, (\bar d_{k R}\gamma^\mu  d_{n R})$
					&
					$    \phantom{+} \lambda^e_{ij} \, \delta_{kn}
								\left[  \frac49 e^2(C_1-3C_3)-4 \, s^2_\theta\, y_t^2 \lambda^u_{33}(C_1-C_3)  \right]$ \\ [3pt]
		\hline
		$(\bar e_{i R}\gamma_\mu e_{j R}) \, (\bar d_{k L}\gamma^\mu  d_{n L})$
					&
					$    \phantom{+} \delta_{ij} \, \lambda^d_{kn}
								\left[  -\frac43 e^2(C_1+C_3)  \right]$ \\ [3pt]
\hline
\end{tabular}
\caption{Operators $Q_i$ and coefficients $\xi_i$ for the semileptonic part of the effective Lagrangian ${\cal L}_{eff}^{EW}$ involving charged leptons and neutral currents.
Generation indices run from 1 to 3, exception made for up-type quarks where $k,n = 1,2$. We set $\sin^2\theta_W\equiv s^2_\theta$.}
\end{table}

\begin{table}[tp] 		
\centering
\begin{tabular}{| c | l |}
		\hline
		$Q_i$& \multicolumn{1}{c |}{$\xi_i$}\\
		\hline
		&\\[-19pt]
		\hline
		$(\bar e_{i L}\gamma_\mu \nu_{j L}) (\bar u_{k L}\gamma^\mu  d_{n L})$
					&
					$   \phantom{+} \lambda^e_{ij} \, \lambda^{ud}_{kn}
								\left[  -6 g_2^2 C_1+2(6g_2^2+g_1^2)C_3  \right]$ \\
					&
					$   + \lambda^e_{ij} \, V^{\mysmall{CKM}}_{kn}
								\left[  -12 \, y_t^2 \lambda^u_{33} C_3  \right]$ \\ 
					&
					$   + \lambda^e_{ij} \, ( \lambda^u_{k3} V^{\mysmall{CKM}}_{3n} +
																			\delta_{k3} \lambda^{ud}_{3n} )
								\left[  - y_t^2 C_3  \right]$ \\ [3pt]
\hline
\end{tabular}
\caption{Operators $Q_i$ and coefficients $\xi_i$ for the semileptonic part of the effective Lagrangian ${\cal L}_{eff}^{EW}$ involving charged currents. 
For up-type quarks the indices rum from 1 to 2. The $\xi_i$ coefficient for the Hermitian conjugate operator can be easily derived. \label{ta:4fCC}}
\end{table}

The main point to stress here is that the purely leptonic contributions to the effective Lagrangian are entirely generated by quantum effects.
They contain terms of order $y_t^2/(16\pi^2)$ and $e^2/(16\pi^2)$. 
In the gauge part, interestingly enough, the surviving contributions are controlled by the electromagnetic coupling $e^2$.
These contributions can thus be interpreted as arising from a photon exchange between the quark loop arising from ${\cal L}_{NP}^0$ and the electromagnetic current.

\subsection{Effective theory below the electroweak breaking scale}
To derive the effective Lagrangian suitable for application to processes at the GeV scale, we should replicate the steps outlined above. From the electroweak scale $m_{EW}$ down to the next mass thresholds,
the bottom  and charm masses $m_b, m_c$, we have to include the further running of the Wilson coefficients. In this regime, the only relevant interaction of the electroweak theory is the electromagnetic one.
At the scale close to $m_b$ ($m_c$), we move to a new effective theory where the $b$ ($c$) quark is integrated out. The new theory is determined from matching conditions analogous to those
described above for the electroweak threshold. We go on until we reach the GeV scale. For simplicity in the following we assume for the light quarks $u$, $d$ and $s$ a common constituent mass 
of order $\sim1$ GeV.

\begin{figure}[tp]
\centering
\includegraphics[width=0.14\textwidth]{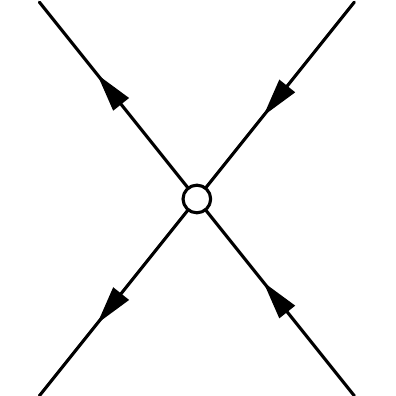}
\includegraphics[width=0.05\textwidth]{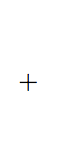}
\includegraphics[width=0.14\textwidth]{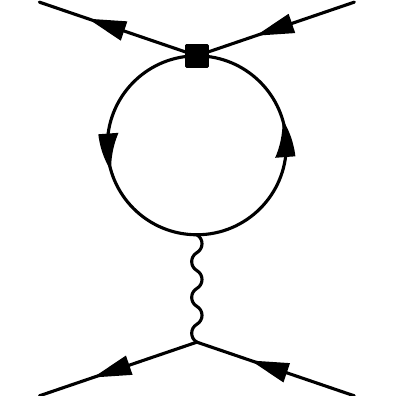}
\includegraphics[width=0.07\textwidth]{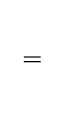}
\includegraphics[width=0.14\textwidth]{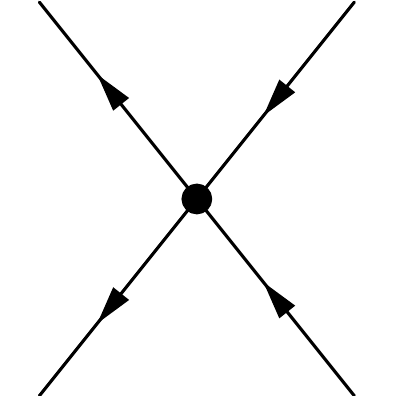}
\includegraphics[width=0.05\textwidth]{plus}
\includegraphics[width=0.14\textwidth]{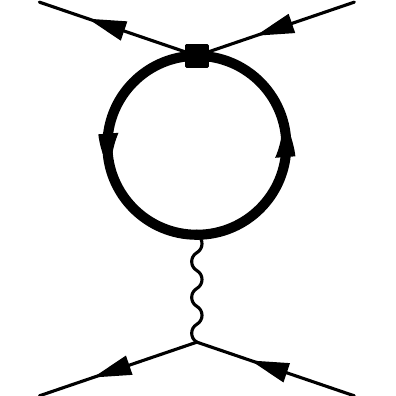}
\includegraphics[width=0.05\textwidth]{plus}
\includegraphics[width=0.14\textwidth]{em2.pdf}
\caption{Matching condition at the $m_b$ ($m_c$) scale for the operator $(\bar e_{iL} \gamma_\mu e_{jL}) \, (\bar e_{kL}\gamma^\mu e_{nL})$. On the right the amplitude is computed above the mass threshold. The thick line represents a bottom (charm) loop. 
On the left the amplitude is evaluated below the mass threshold. The tree-level diagram with an open circle is the unknown. The loop diagrams with thin lines cancel among each other.}
\label{em}
\end{figure}


Again, we will exemplify the matching procedure by performing it explicitly for the operator $(\bar e_{iL} \gamma_\mu e_{jL}) \, (\bar e_{kL}\gamma^\mu e_{nL})$.
The corresponding matching condition is shown in fig. \ref{em}. On the right-hand side the process is evaluated above $m_b$ ($m_c$), where the thick line represents a bottom (charm) loop. On the left hand side the same process is computed below the $m_b$ ($m_c$) scale where the bottom (charm) loop is absent. After crossing both mass thresholds
we get
\begin{align}
\cA(\mu) = \frac{\lambda^e_{ij} \, \delta_{kn} }{16\pi^2\Lambda^2} 
						\bigg\{ \!&
						-12(-\frac{1}{2}+\sin^2\theta_W)y_t^2 \lambda^u_{33}(C_1-C_3) \log\frac{\Lambda}{m_t}
						+ \frac43 e^2 (C_1-3C_3) \log\frac{\Lambda}{\mu}
						\nn \\
						&
						+
							  \frac43 e^2
								\left[ - 2(C_1-C_3) (\lambda^u_{33}\log\frac{m_t}{\mu}
																									+\lambda^u_{22}\log\frac{m_c}{\mu})\right. 
										\left. +(C_1+C_3) \lambda^d_{33}\log\frac{m_b}{\mu}  \right] \bigg\}~.
\end{align}
This is the Wilson coefficient for $(\bar e_{iL} \gamma_\mu e_{jL}) \, (\bar e_{kL}\gamma^\mu e_{nL})$, at a scale $\mu$ below the charm threshold. 

In a similar way, the Wilson coefficient for the full effective Lagrangian at the GeV scale can be derived. 
We recast the final result in the following form:
\be
{\cal L}_{eff}={\cal L}_{eff}^{EW}+\delta{\cal L}_{eff}^{QED} ~,
\ee
where $\cL_{eff}^{EW}$ is given in eq.~(\ref{LEWlow}), while
\be
\delta{\cal L}_{eff}^{QED}=\frac{1}{16\pi^2\Lambda^2}\log\frac{m_{EW}}{\mu}~\sum_i \delta\xi_i~ Q_i ~.
\ee
The operators $Q_i$ and the coefficients $\delta\xi_i$ for the four-fermion operators of purely leptonic and semileptonic type 
are given in Table \ref{tab:QEDlep}--\ref{tab:QEDcc}.

When computing the amplitude for a physical process, two contributions are generally required: the first is formally a tree-level amplitude described by the RGE-induced part of our effective Lagrangian. The second is a genuine
one-loop amplitude where the residual light degrees of freedom of the low-energy theory are exchanged in the internal lines. These two terms correspond, for instance, to the two diagrams drawn in the left-hand side of Fig. 4.
We will keep only the leading logarithmic part of each term. They are both scale dependent, but this dependence cancels in the sum. The neglected finite contributions do not depend on the scale $\mu$. To deal with loops with the light virtual quarks $u$, $d$ and $s$, we will assume a common constituent 
mass $\hat\mu\approx$ 1 GeV.
\begin{table}[t] 		
\centering
\begin{tabular}{| c | l |}
		\hline
		\multicolumn{1}{| c |}{$Q_i$} & \multicolumn{1}{c |}{$\delta\xi_i$}\\
		\hline
		&\\[-19pt]
		\hline
		$(\bar\nu_{i L}\gamma_\mu \nu_{j L})~(\bar \nu_{k L}\gamma^\mu  \nu_{n L})$
					&
					$ \phantom{+} 0$ \\ [3pt]
		\hline
		$(\bar\nu_{i L}\gamma_\mu \nu_{j L})~(\bar e_{k}\gamma^\mu  e_{n})$
					&
					$ \phantom{+} \lambda^e_{ij} \, \delta_{kn}
								\cdot  \frac43 e^2
								\left[  (C_1+3C_3) - 2(C_1+C_3) (\hat{\lambda}^u_{33}\log\frac{m_t}{\mu}
																									+\hat{\lambda}^u_{22}\log\frac{m_c}{\mu})\right.$ \\
								&
										$ \phantom{+ \lambda^e_{ij} \, \delta_{kn} \cdot  \frac43 e^2 [}
										\left. +(C_1-C_3)\hat{\lambda}^d_{33}\log\frac{m_b}{\mu}  \right]$ \\ [5pt]
		\hline
		$(\bar e_{i L}\gamma_\mu e_{j L})~(\bar e_{k}\gamma^\mu  e_{n})$
					&
					$ \phantom{+} \lambda^e_{ij} \, \delta_{kn}
								\cdot  \frac43 e^2
								\left[  (C_1-3C_3) - 2(C_1-C_3) (\hat{\lambda}^u_{33}\log\frac{m_t}{\mu}
																									+\hat{\lambda}^u_{22}\log\frac{m_c}{\mu})\right.$ \\
								&
										$ \phantom{+ \lambda^e_{ij} \, \delta_{kn} \cdot  \frac43 e^2 [}
										\left. +(C_1+C_3)\hat{\lambda}^d_{33}\log\frac{m_b}{\mu}  \right]$ \\ [5pt]
\hline
\end{tabular}
\caption{Operators $Q_i$ and coefficients $\delta\xi_i$ for the purely leptonic part of the effective Lagrangian $\delta{\cal L}_{eff}^{QED}$. We set $\hat{\lambda}^{u,d}_{ii}=\lambda^{u,d}_{ii}/\log\frac{m_{EW}}{\mu}$.
\label{tab:QEDlep}}
\end{table}

\begin{table}[t] 		
\centering
\begin{tabular}{| c | l |}
		\hline
		$Q_i$& \multicolumn{1}{c |}{$\delta\xi_i$}\\
		\hline
		&\\[-19pt]
		\hline
		$(\bar\nu_{i L}\gamma_\mu \nu_{j L}) \, (\bar u_{k}\gamma^\mu  u_{n})$
					&
					$ \phantom{+}   \lambda^e_{ij} \, \delta_{kn}  \left(-\frac89 e^2 \right)
								\left[   (C_1+3C_3)-2(C_1+C_3)(\hat{\lambda}^u_{33}\log\frac{m_t}{\mu}+\hat{\lambda}^u_{22}\log\frac{m_c}{\mu})  \right.$ \\
								 &
								 $ \phantom{+ \lambda^e_{ij} \, \delta_{kn}  \left(-\frac89 e^2 \right) [ }
								 \left. +(C_1-C_3)\hat{\lambda}^d_{33}\log\frac{m_b}{\mu} \right]$ \\ [5pt]
		\hline
		$(\bar\nu_{i L}\gamma_\mu \nu_{j L}) \, (\bar d_{k}\gamma^\mu  d_{n})$
					&
					$ \phantom{+}   \lambda^e_{ij} \, \delta_{kn}  \cdot \frac49 e^2 
								\left[   (C_1+3C_3)-2(C_1+C_3)(\hat{\lambda}^u_{33}\log\frac{m_t}{\mu}+\hat{\lambda}^u_{22}\log\frac{m_c}{\mu})  \right.$ \\
								 &
								 $ \phantom{+ \lambda^e_{ij} \, \delta_{kn}  \cdot \frac49 e^2 [ }
								 \left. +(C_1-C_3)\hat{\lambda}^d_{33}\log\frac{m_b}{\mu} \right]$ \\ [5pt]
\hline
\end{tabular}
\caption{Operators $Q_i$ and coefficients $\delta\xi_i$ for the semileptonic part of the effective Lagrangian $\delta{\cal L}_{eff}^{QED}$ involving neutrinos and neutral currents.
For the down-type quarks generation indices run from 1 to 2,while for up-type quarks we only keep the first generation. We set $\hat{\lambda}^{u,d}_{ii}=\lambda^{u,d}_{ii}/\log\frac{m_{EW}}{\mu}$.}
\end{table}

\begin{table}[hp] 		
\centering
\begin{tabular}{| c | l |}
		\hline
		$Q_i$& \multicolumn{1}{c |}{$\delta\xi_i$}\\
		\hline
		&\\[-19pt]
		\hline
		$(\bar e_{i L}\gamma_\mu e_{j L}) \, (\bar u_{k L}\gamma^\mu  u_{n L})$
					&
					$    \phantom{+} \lambda^e_{ij} \, \lambda^u_{kn} 
								\left[ 8 e^2 \,(C_1-C_3) \right]$ \\ 
					&
					$    + \delta_{ij} \, \lambda^u_{kn}
								\left[  -\frac{4}{3} e^2(C_1-C_3)  \right]$ \\ 
					&
					$    - \lambda^e_{ij} \, \delta_{kn}  \cdot
								 \frac89 e^2\left[(C_1-3C_3)-2(C_1-C_3)(\hat{\lambda}^u_{33}\log\frac{m_t}{\mu}
																							+\hat{\lambda}^u_{22}\log\frac{m_c}{\mu})\right.$ \\
								&
								$ \phantom{ - \lambda^e_{ij} \, \delta_{kn}  \cdot \frac89 e^2 [ }
								\left.+(C_1+C_3)\hat{\lambda}^d_{33}\log\frac{m_b}{\mu}  \right]$ \\ [3pt]
		\hline
		$(\bar e_{i L}\gamma_\mu e_{j L}) \, (\bar u_{k R}\gamma^\mu  u_{n R})$
					&
					$   - \lambda^e_{ij} \, \delta_{kn} \cdot
								 \frac89 e^2\left[(C_1-3C_3)-2(C_1-C_3)(\hat{\lambda}^u_{33}\log\frac{m_t}{\mu}
																							+\hat{\lambda}^u_{22}\log\frac{m_c}{\mu})\right.$ \\
								&
								$ \phantom{ - \lambda^e_{ij} \, \delta_{kn}  \cdot \frac89 e^2 [ }
								\left.+(C_1+C_3)\hat{\lambda}^d_{33}\log\frac{m_b}{\mu}  \right]$ \\ [3pt]
		\hline
		$(\bar e_{i R}\gamma_\mu e_{j R}) \, (\bar u_{k L}\gamma^\mu  u_{n L})$
					&
					$    \phantom{+} \delta_{ij} \, \lambda^u_{kn}
								\left[  -\frac{4}{3} e^2(C_1-C_3)  \right]$ \\ [3pt]
		\hline
		$(\bar e_{i L}\gamma_\mu e_{j L}) \, (\bar d_{k L}\gamma^\mu  d_{n L})$
					&
					$    \phantom{+} \lambda^e_{ij} \, \lambda^d_{kn}
								\left[  -4 e^2 \,(C_1+C_3)  \right]$ \\ 
					&
					$    + \delta_{ij} \, \lambda^d_{kn}
								\left[  -\frac43 e^2(C_1+C_3)  \right]$ \\ 
					&

					$    + \lambda^e_{ij} \, \delta_{kn} \cdot \frac49 e^2
								\left[(C_1-3C_3)-2(C_1-C_3)(\hat{\lambda}^u_{33}\log\frac{m_t}{\mu}
																							+\hat{\lambda}^u_{22}\log\frac{m_c}{\mu})\right.$ \\
								&
								$ \phantom{ - \lambda^e_{ij} \, \delta_{kn}  \cdot \frac89 e^2 [ }
								\left.+(C_1+C_3)\hat{\lambda}^d_{33}\log\frac{m_b}{\mu}  \right]$ \\ [3pt]
		\hline
		$(\bar e_{i L}\gamma_\mu e_{j L}) \, (\bar d_{k R}\gamma^\mu  d_{n R})$
					&
					$    \phantom{+} \lambda^e_{ij} \, \delta_{kn} \cdot \frac49 e^2
								\left[(C_1-3C_3)-2(C_1-C_3)(\hat{\lambda}^u_{33}\log\frac{m_t}{\mu}
																							+\hat{\lambda}^u_{22}\log\frac{m_c}{\mu})\right.$ \\
								&
								$ \phantom{ - \lambda^e_{ij} \, \delta_{kn}  \cdot \frac89 e^2 [ }
								\left.+(C_1+C_3)\hat{\lambda}^d_{33}\log\frac{m_b}{\mu}  \right]$ \\ [3pt]
		\hline
		$(\bar e_{i R}\gamma_\mu e_{j R}) \, (\bar d_{k L}\gamma^\mu  d_{n L})$
					&
					$    \phantom{+} \delta_{ij} \, \lambda^d_{kn}
								\left[  -\frac43 e^2(C_1+C_3)  \right]$ \\ [3pt]
\hline
\end{tabular}
\caption{Operators $Q_i$ and coefficients $\delta\xi_i$ for the semileptonic part of the effective Lagrangian $\delta{\cal L}_{eff}^{QED}$ involving charged leptons and neutral currents.
For the down-type quarks generation indices run from 1 to 2, while for up-type quarks we only keep the first generation. We set  $\hat{\lambda}^{u,d}_{ii}=\lambda^{u,d}_{ii}/\log\frac{m_{EW}}{\mu}$.}
\end{table}

\begin{table}[hp] 		
\centering
\begin{tabular}{| c | l |}
		\hline
		$Q_i$& \multicolumn{1}{c |}{$\delta\xi_i$}\\
		\hline
		&\\[-19pt]
		\hline
		$(\bar e_{i L}\gamma_\mu \nu_{j L}) (\bar u_{k L}\gamma^\mu  d_{n L})$
					&
					$   -  \lambda^e_{ij} \, \lambda^{ud}_{kn} \cdot 8 e^2 \, C_3$ \\ [3pt]
\hline
\end{tabular}
\caption{Operators $Q_i$ and coefficients $\delta\xi_i$ for the semileptonic part of the effective Lagrangian $\delta{\cal L}_{eff}^{QED}$ involving charged currents. 
For the down-type quarks generation indices run from 1 to 2, while for up-type quarks we only keep the first generation. The $\delta\xi_i$ coefficient for the Hermitian conjugate operator can be easily derived.
\label{tab:QEDcc}}
\end{table}

\section{Observables}
\label{sec_3}
In this section we analyze the phenomenological implications of Lagrangian (\ref{LNP}), making use of the RGE-improved low-energy 
EFT derived extensively in the previous section. 

As discussed in section \ref{sec_massbasis}, the full set of free parameters in our set-up are $C_{1,3}/\Lambda^2$ and the two matrices $\lambda^{e,d}$.
In addition, we will assume $\lambda^{e,d}_{11}=\lambda^{e,d}_{12} = \lambda^{e,d}_{13} =0$. This will provide us with a simpler and 
yet conservative framework, since LFUV effects (i.e.~$R_{D^{(\ast)}}$ and $R_{K^{(*)}}$ anomalies) can be easily shown to be 
maximized by such assumption. In conclusion, our setup contains effectively only four\footnote{More precisely, $\Lambda$ should be considered as a fifth independent parameter, governing the size of RGE effects. Since these effects depend on $\Lambda$, which is of order TeV, only very mildly, we can approximately regard $\Lambda$ as a fixed quantity in the relevant logarithms.} free parameters:
\begin{align*}
C_{1,3}/\Lambda^2 ~,  \qquad \lambda^{e,d}_{23} ~,
 \end{align*} 
while all other non-vanishing matrix elements can be derived through eq.~(\ref{parametrization}) and the property ${\rm tr}\lambda^f=1$. Furthermore, we will assume $\lambda^{e,d}_{22} \approx |\lambda^{e,d}_{23}|^2 \ll \lambda^{e,d}_{33}$ and also $\lambda^{e,d}_{23}$ to 
be real in the CKM basis. Given eq.~(\ref{parametrization}) and our assumptions, we straightforwardly derive $|\lambda^{e,d}_{23}|\leq 1/2$.

Rather than aiming at a complete investigation of the wide phenomenology triggered by the Lagrangian (\ref{LNP}), we prefer to focus 
our analysis on processes that, despite the loop suppression, can compete with tree-level semileptonic bounds thanks to their high experimental resolutions. Among them, arguably the most interesting ones are the fully leptonic processes and the leptonic decays of the $Z$ vector boson.
We structure the present section as follows.  
In section 3.1, we discuss how to address both charged- and neutral-current $B$ anomalies within our framework. 
In section 3.2, we discuss the most relevant tree-level phenomenology connected with the $B$ anomalies. 
In section 3.3, we proceed to study observables in the leptonic sector receiving large contributions at loop-level.
In section 3.4, a global numerical analysis is performed in order to corroborate our main message, namely that bounds 
coming from one-loop induced lepton phenomenology play a major role while trying to address the $B$ anomalies, and particularly $R_{D^{(\ast)}}$.

\subsection{The \texorpdfstring{$B$}{B} anomalies}
Here we recall how the $B$-anomalies can be simultaneously explained by extending the SM through the addition of ${\cal L}_{NP}^0(\Lambda)$ in eq. (\ref{LNP}).
To this purpose NP should contribute dominantly to charged-current transitions compared to the neutral-current ones, since in the SM the former arise at the tree-level and the latter at one-loop.
This can be reproduced within the present framework by assuming an hierarchy between $\lambda^d_{33}\lambda^e_{33}$, which controls $B \to D^{(*)} \tau \nu$, and $\lambda^d_{23}\lambda^e_{22}$, which is responsible for
$B\to K\mu^+\mu^-$.

\subsubsection{\texorpdfstring{$B\to K\ell\bar\ell$}{B to Kll}}
%
In our setup, the leading $SU(3)_c \times U(1)_{el}$ invariant effective Lagrangian describing the semileptonic 
process $b \to s \,\bar e_i e_j$ is~\cite{Buchalla:1995vs}
\begin{equation}
  {\cal{L}}^{\mysmall \rm NC}_{\rm eff}=
  \frac{4 G_F}{\sqrt{2}}  \lambda_{bs}  \left( C^{ij}_9  \mathcal{O}^{ij}_9  +  C^{ij}_{10}  \mathcal{O}^{ij}_{10}  \right)+h.c.\,,
\end{equation}
where $\lambda_{bs} \!=\! V_{tb}^{} V_{ts}^\ast$ and the operators $\mathcal{O}_{9,10}$ read
\begin{align}
  \mathcal{O}^{ij}_{9} & = \frac{e^2}{(4 \pi)^2} (\bar{s}_L \gamma_\mu b_L)(\bar{e}_i \gamma^\mu e_j)\,,\\
  \mathcal{O}^{ij}_{10}  &= \frac{e^2}{(4 \pi)^2} (\bar{s}_L \gamma_\mu b_L)(\bar{e}_i \gamma^\mu \gamma_5 e_j)\,.
\end{align}
By matching ${\cal{L}}^{\mysmall \rm NC}_{\rm eff}$ with ${\cal L}_{\rm NP}^0$ at the tree level, we obtain:
\begin{align}
(C^{\rm\mysmall NP}_9)^{ij} = &-(C^{\rm\mysmall NP}_{10})^{ij} 
= \frac{\pi}{\alpha\lambda_{bs}}\frac{v^2}{\Lambda^2}\,(C_{1} + C_{3})\,\lambda^d_{23}\lambda^e_{ij} + \cdots\,,
\end{align}
where dots stand for subleading RGE induced terms,  typically negligible compared to the leading ones.

As shown in eqs.~\ref{eq:RKSexp} and \ref{eq:RKexp}, the experimental values of $\RKss$, which accounts 
for LFUV in the process $B\to K^{(*)} \ell\bar\ell$, read
\ba
\RK =  0. 745 {}^{+0.090}_{-0.074} \pm 0.036~,
\qquad\quad
\RKs = 0. 685 {}^{+0.113}_{-0.069} \pm 0.047~.
\label{eq:R_K_exp}
\ea
In our framework, $\RK$ is well approximated by the expression
\begin{equation}
\RK \approx  
\frac{|(C_9^{\mysmall\rm NP})^{22} + C_9^{\mysmall\rm SM}|^2 + |(C_{10}^{\mysmall\rm NP})^{22} + C_{10}^{\mysmall\rm SM}|^2}
{|(C_9^{\mysmall\rm NP})^{11} + C_9^{\mysmall\rm SM}|^2 + |(C_{10}^{\mysmall\rm NP})^{11} + C_{10}^{\mysmall\rm SM}|^2}\,.
\label{eq:R_K}
\end{equation}
Given $C^{\mysmall\rm SM}_9 \approx - C^{\mysmall\rm SM}_{10} \approx 4.2$, assuming
$C^{\mysmall\rm NP}_9 \approx - C^{\mysmall\rm NP}_{10}$ and $(C_9^{\mysmall\rm NP})^{11} = 0$, and working in a linear approximation for the NP contribution, we obtain
\begin{equation}
\RK \approx  1 - \frac{2\pi}{\alpha |V_{ts}||C_9^{\mysmall\rm SM}|}\frac{v^2}{\Lambda^2} (C_1 + C_3) \lambda^d_{23}\lambda^e_{22}
\approx  1 - 0.28 \, \frac{(C_1 + C_3)}{\Lambda^2({\rm TeV}^2)} \frac{\lambda^d_{23}\,\lambda^e_{22}}{10^{-3}}\,.
\label{an1}
\end{equation}
As a result, for $\Lambda = 1$ TeV, $\RK$ can be accounted for by $\mathcal{O}(1)$ values of $C_1+C_3$ 
and flavour mixing angles of order $\lambda^d_{23} \lambda^e_{22} \sim 10^{-3}$.

On the other hand, Eq.~(\ref{eq:R_K}) cannot be directly applied to $\RKs$, since in that case a non-trivial role is played by the 
Wilson coefficient $C_7$~\cite{Altmannshofer:2008dz}. 
Nevertheless, under our NP assumptions the central value for $\RKs$ differs less than $10\%$ from the $\RK$ value given by 
(\ref{eq:R_K}), 
thus not affecting our semi-quantitative arguments.

\subsubsection{\texorpdfstring{$B \to D^{(*)} \ell \nu$}{B to Dln}}

The effective Lagrangian relevant for charged-current processes like $b\to c\ell\nu$ is given by
\begin{equation}
\!\!\!{\cal{L}}^{\mysmall \rm CC}_{\rm eff} \!=\! 
-\frac{4 G_F}{\sqrt{2}} \, V_{cb} \, (C^{cb}_L)_{ij}\left( \bar c_L \gamma_\mu b_L  \right) \left( \bar e_{Li} \gamma^\mu \nu_{Lj}  \right) + h.c.\,.
\end{equation}
In our model, for negligible values of $\lambda^{d}_{13}$, the coefficient $(C^{cb}_L)_{ij}$ reads
\begin{equation}
(C^{cb}_L)_{ij} \approx \delta_{ij} - \frac{v^2}{\Lambda^2} \,C_3\, \lambda^{e}_{ij}  \frac{\lambda^{ud}_{23}}{V_{cb}}
\approx
\delta_{ij} - \frac{v^2}{\Lambda^2} \,  C_3 \, \lambda^e_{ij} \left( \frac{V_{cs}}{V_{cb}} \lambda^d_{23} + \lambda^d_{33}\right)
\,.
\end{equation}
LFUV in the charged-current process $B \to D^{(*)} \ell \nu$ is encoded in the observable $R^{\tau/\ell}_{D^{(*)}}$,
see eqs. \ref{eq:RDexp},\ref{eq:RDSexp}:
\ba
R^{\tau/\ell}_{D^*} = 1.23 \pm 0.07~,  \qquad\qquad R^{\tau/\ell}_{D } = 1.34 \pm 0.17~.
\ea
In our framework, $R^{\tau/\ell}_{D^{(*)}}$ reads
\begin{equation}
R^{\tau/\ell}_{D^{(*)}} = \frac{\sum_{j} |(C^{cb}_{L})_{3j}|^2}{\sum_{j} |(C^{cb}_{L})_{\ell j}|^2}\,,
\end{equation}
and therefore, for $\lambda^{e}_{22} \!\ll\! \lambda^{e}_{33}$, we obtain
\begin{equation}
R^{\tau/\ell}_{D^{(*)}} \approx 1 - \frac{2v^2}{\Lambda^2} \,  C_3~ \lambda^{e}_{33}\, \left( \frac{V_{cs}}{V_{cb}} \lambda^d_{23} + \lambda^d_{33}\right)
\approx 1 - \frac{0.12 \, C_3}{\Lambda^{2}(\rm TeV^2)} \left( \frac{V_{cs}}{V_{cb}} \lambda^d_{23} + \lambda^d_{33}\right)\,,
\label{an2}
\end{equation}
where we took $\lambda^e_{33}\approx 1$. As a result, in order to accommodate the $R^{\tau/\ell}_{D^{(*)}}$ anomaly, we need $C_3<0$ 
and $C_3 \sim \mathcal{O}(1)$, for $\Lambda=1$ TeV.
The condition $\lambda^{e}_{22} \!\ll\! \lambda^{e}_{33}$ is justified a posteriori by the non observation of LFUV in the $\mu/e$ sector up to 
the $\lesssim 2\%$ level~\cite{Agashe:2014kda,Greljo:2015mma}. Indeed, from the expression of $R^{\mu/e}_{D^{(*)}}$
\begin{equation}
R^{\mu/e}_{D^{(*)}} \approx 1 - \frac{2v^2}{\Lambda^2} \,  C_3 \,\lambda^e_{22} \left( \frac{V_{cs}}{V_{cb}} \lambda^d_{23} + \lambda^d_{33}\right)
\approx 1 - \frac{0.01 \, C_3}{\Lambda^{2}(\rm TeV^2)} 
\left(\frac{\lambda^e_{22}}{0.1}\right)
\left( \frac{V_{cs}}{V_{cb}} \lambda^d_{23} + \lambda^d_{33}\right)\,,
\end{equation}
we find the upper bound $\lambda^e_{22} \lesssim 0.1$ once the anomaly in the $\tau/\ell$ sector is explained. 
Notice that in our estimates we always set $\lambda^d_{11}=0$, as well as $\lambda^e_{11}=0$ which implies 
$\lambda^e_{22} \approx (\lambda^{e}_{23})^2$.

%
\subsection{Tree-level semileptonic phenomenology}
An immediate consequence of the adopted framework is a set of deviations predicted in leptonic and semileptonic $B$-decays strictly
related to the anomalous channels discussed in the previous section. The modifications with respect to the SM predictions are entirely dominated by the Lagrangian ${\cal L}_{NP}^0(\Lambda)$
and do not need the inclusion of quantum effects, at least in generic regions of the parameter space. 
In our framework $\lambda^e_{22}$, $\lambda^e_{33}$ and thus $\lambda^e_{23}$ are non vanishing, implying
LFV in B meson decays.
The processes presented in this section have been widely discussed in the literature.
We list them here for completeness and we analyze the bounds on the parameters they give rise to.

\subsubsection{\texorpdfstring{$B \to \ell\nu$}{B to ln}}
%
In our framework, another charged-current process which is closely related to $B \to D^{(*)}\ell\nu$ is the decay
$B \to \ell\nu$. The related LFUV observable, $R^{\tau/\ell}_{B\tau\nu}$, is defined as
\begin{align}
R^{\tau/\mu}_{B\tau\nu} = 
\frac{\mathcal{B}(B \to \tau\nu)_{\rm exp}/\mathcal{B}(B \to \tau\nu)_{\rm SM}}{\mathcal{B}(B \to \mu\nu)_{\rm exp}/\mathcal{B}(B \to \mu\nu)_{\rm SM}} 
\,,
\end{align}
which can be evaluated by means of the expression
\begin{equation}
R^{\tau/\mu}_{B\tau\nu} = \frac{\sum_{j} |(C^{ub}_{L})_{3j}|^2}{\sum_{j} |(C^{ub}_{L})_{2 j}|^2}\,,
\end{equation}
where
\begin{equation}
(C^{ub}_L)_{ij} \approx \delta_{ij} - \frac{v^2}{\Lambda^2} \,C_3\, \lambda^{e}_{ij} \, \frac{\lambda^{ud}_{13}}{V_{ub}}
\approx
\delta_{ij} - \frac{v^2}{\Lambda^2} \,  C_3 \, \lambda^e_{ij} \left( \lambda^d_{33} + \frac{V_{us}}{V_{ub}} \lambda^d_{23} \right)
\,.
\end{equation}
Assuming that $\lambda^d_{23} \ll \lambda^d_{33} \simeq 1$ and $\lambda^e_{22} \ll \lambda^e_{33} \simeq 1$, 
the expression for $R^{\tau/\ell}_{D^{(*)}}$ reads
\begin{equation}
R^{\tau/\mu}_{B\tau\nu} \approx 1 - \frac{2v^2}{\Lambda^2} \,  C_3 \,
\left( 1 + \frac{V_{us}\lambda^d_{23}}{|V_{ub}|\cos\gamma} \right)
\,,
\end{equation}
where $\gamma \approx 70^\circ$.
Since Belle II aims to measure $R^{\tau/\mu}_{B\tau\nu}$ with a $5\%$ resolution, it is likely that 
$R^{\tau/\mu}_{B\tau\nu}$ will provide a strong constraint to the present framework.

\subsubsection{\texorpdfstring{$B\to K^{(*)} \nu\bar\nu$}{B to Knn}}
%
The leading $SU(3)_c \times U(1)_{el}$ invariant effective Lagrangian describing the semileptonic 
process $b \to s \,\bar \nu_i \nu_j$ is~\cite{Buchalla:1995vs} is given in our framework by
\begin{equation}
  {\cal{L}}^{\mysmall \rm NC}_{\rm eff} =  \frac{4 G_F}{\sqrt{2}}  \lambda_{bs} C^{ij}_\nu  \mathcal{O}^{ij}_\nu  +h.c.\,,
\end{equation}
where the operator $\mathcal{O}^{ij}_{\nu}$ reads
\begin{align}
  \mathcal{O}^{ij}_{\nu}  &= 
    \frac{e^2}{(4 \pi)^2} (\bar{s}_L \gamma_\mu b_L)(\bar{\nu}_i \gamma^\mu (1\!-\!\gamma_5) \nu_j)\,.
\end{align}
By matching ${\cal{L}}^{\mysmall \rm NC}_{\rm eff}$ with ${\cal L}_{\rm NP}$, we obtain:
\begin{align}
(C^{\rm\mysmall NP}_\nu)^{ij} = &\frac{\pi}{\alpha\lambda_{bs}}\frac{v^2}{\Lambda^2}
\,(C_{1} \!-\! C_{3})\,
\lambda^d_{23}\lambda^e_{ij} + \cdots \,,
\end{align}
where, again, dots stand for RGE induced terms which are always subleading, unless $C_1 = C_3$. 
Interestingly, the latter condition can be realised in scenarios with vector leptoquark mediators~\cite{Calibbi:2015kma}.
In such a case, the RGE induced effects to $(C^{\rm\mysmall NP}_\nu)^{ij}$ are given by:
\begin{align}
(\delta C^{\rm\mysmall NP}_\nu)^{ij}
\simeq -\frac{3g^2_2}{4\pi\alpha\lambda_{bs}}\frac{v^2}{\Lambda^2}
C_3 \log\frac{\Lambda}{m_{EW}} \,
\lambda^d_{23}\lambda^e_{ij} 
\,,
\end{align}
which are of order $|(\delta C^{\rm\mysmall NP}_\nu)^{ij}| /  |(C^{\rm\mysmall NP}_\nu)^{ij}| \approx 0.1 \times C_3/(C_3-C_1)$ for $\Lambda=1~$TeV.
As already observed in~\cite{Calibbi:2015kma}, the process $B\to K^{(*)} \nu\bar\nu$ sets relevant constraints on our model.
Defining $R^{\nu\nu}_{K^{(*)}}$ as
\begin{equation}
R^{\nu\nu}_{K^{(*)}} = \frac{\mathcal{B}(B\to K^{(*)} \nu\bar\nu)}{\mathcal{B}(B\to K^{(*)} \nu\bar\nu)_{\mysmall\rm SM}}
\,,
\end{equation}
we get
\begin{equation}
R^{\nu\nu}_{K^{(*)}} = \frac{\sum_{ij} 
|C^{\mysmall \rm SM}_\nu \,\delta^{ij}+ (C^{\rm\mysmall NP}_\nu)^{ij} |^2}{3|C^{\mysmall \rm SM}_\nu |^2}\,,
\end{equation}
where $C^{\mysmall \rm SM}_\nu \approx -6.4$ so that $R^{\nu\nu}_{K^{(*)}}$ reads
\begin{equation}
R^{\nu\nu}_{K^{(*)}} \simeq 1 + \frac23   \left(\frac{\pi}{\alpha |V_{ts}|}\frac{v^2}{\Lambda^2} \frac{(C_1-C_3)}{|C^{\mysmall \rm SM}_\nu|}\right)
\lambda^d_{23} 
+ \frac13  \left( \frac{\pi}{\alpha |V_{ts}|}\frac{v^2}{\Lambda^2} \frac{(C_1-C_3)}{|C^{\mysmall \rm SM}_\nu|} \right)^{\!2} \!(\lambda^d_{23})^2
\,.
\end{equation}
The above expression has been obtained setting $\lambda_{bs}\simeq - |V_{ts}|$ and using
the properties ${\rm tr}\lambda^f = 1$ and  $\sum_{ij} |\lambda^f_{ij}|^2 = 1$.
Therefore, $R^{\nu\nu}_{K^{(*)}}$ is well approximated by the numerical expression
\begin{equation}
R^{\nu\nu}_{K^{(*)}} \approx 1 + 0.6 \, \frac{(C_1-C_3)}{\Lambda^{2} {(\rm TeV^2)}} 
\left(\frac{\lambda^d_{23}}{0.01} \right)
+ 0.3 \, \frac{(C_1-C_3)^{2}}{\Lambda^{4} {(\rm TeV^4)}} 
\left(\frac{\lambda^d_{23}}{0.01} \right)^2
\,,
\end{equation}
showing that for natural values of $C_1-C_3 \sim \mathcal{O}(1)$, $\lambda^d_{23}\sim 10^{-2}$ and $\Lambda=1~$TeV,
$R^{\nu\nu}_{K^{(*)}}$ can easily satisfy the experimental bounds
\begin{equation}
R^{\nu\nu}_{K} <  4.3 \qquad\qquad\qquad R^{\nu\nu}_{K^*} <  4.4\,.
\end{equation}

\subsubsection{\texorpdfstring{$B_{s}\to\mu\bar\mu$}{Bs to mm}}

Since our model predicts the relation $C_9=-C_{10}$, an explanation of the $R^{\mu/e}_K$ anomaly implies
NP contributions also for the decay mode $B_{s}\to\mu\bar\mu$. The current experimental measurement and SM
prediction for the branching ratio of this process are~\cite{CMS:2014xfa,Bobeth:2013uxa}:
\begin{equation}
\mathcal{B}(B_s\to\mu\bar\mu)_{\rm exp}=2.8^{+0.7}_{-0.6}\times10^{-9}\qquad\qquad
\mathcal{B}(B_s\to\mu\bar\mu)_{\rm SM}= 3.65(23)\times10^{-9}\,.
\end{equation}
Defining $R_{B_s\mu\mu}$ as
\begin{equation}
R_{B_s\mu\mu} = \frac{\mathcal{B}(B_s\to\mu\bar\mu)_{\rm exp}}{\mathcal{B}(B_s\to\mu\bar\mu)_{\rm SM}} \simeq 
\left|\frac{C^{\rm SM}_{10} + (C^{\rm NP}_{10})^{22}}{C_{10}^{\rm SM}}\right|^2, 
\label{eq:R}
\end{equation}
where $C_{10}^{\rm SM}\approx -4.2$ and remembering that in our framework $R^{\mu/e}_K$ favours $(C^{\rm NP}_{10})^{22} \approx 0.5$,
we see that an explanation of the $R^{\mu/e}_K$ anomaly improves the agreement with the 
$B_{s}\to\mu\bar\mu$ data.

\subsubsection{Lepton-flavour violating \texorpdfstring{\boldmath$B$\unboldmath}{B}\ decays}
\label{susec:H}

In our setting, LFV B-decays such as $B_s \to \tau^\pm \mu^\mp$ and $B\to K^{(*)} \tau^\pm \mu^\mp$
arise already at the tree level.
Here we give the expressions for their branching ratios assuming $C_9 = -C_{10}$~\cite{Crivellin:2015era}
\begin{align}
\mathcal{B}\left(B_s \to \ell^\pm \ell^{\prime\mp} \right) &\simeq 4\times 10^{-8}\left| C_{9}^{\ell\ell^{\prime}}\right|^2
\nonumber\\
\mathcal{B}(B\to K \ell^\pm \ell^{\prime\mp}) &\simeq 2 \times 10^{-9} 
\left(a_{K\ell\ell^\prime} + b_{K\ell\ell^\prime} \right) \left| C_{9}^{\ell\ell^{\prime}}\right|^2
\nonumber\\
\mathcal{B}(B\to K^{*} \ell^\pm \ell^{\prime\mp}) &\simeq 2 \times 10^{-9} 
\left(a_{K^{*}\ell\ell^\prime} + b_{K^{*}\ell\ell^\prime} + c_{K^*\ell\ell^\prime} + d_{K^*\ell\ell^\prime} \right)
\left| C_{9}^{\ell\ell^{\prime}}\right|^2
\,,
\label{bkstaumu}
\end{align}
where the coefficients $a_i$, $b_i$ $c_i$ and  $d_i$ read
\begin{center}
\begin{tabular}{|c|c|c|c|c|c|c|}
\hline
$\ell\ell^\prime $ & $a_{K\ell\ell^\prime}$ & $b_{K\ell\ell^\prime}$ &
$a_{K^*\ell\ell^\prime}$ & $b_{K^*\ell\ell^\prime}$ &
$c_{K^*\ell\ell^\prime}$ & $d_{K^*\ell\ell^\prime}$ \\
\hline
$\;\tau\mu,\tau e\;$ & $\;9.6 \pm 1.0\;$ & $\;10.0 \pm 1.3\;$ & $\;3.0 \pm
0.8\;$ & $\;2.7 \pm 0.7\;$ & $\;16.4 \pm 2.1\;$ & $\;15.4 \pm 1.9\;$
\\
\hline
\end{tabular}
\,.
\end{center}
We observe that the above branching ratios have been multiplied by a factor of two since the experimental bounds 
refer to the final state $\ell^\pm \ell^{\prime\mp} = \ell^+ \ell^{\prime -} + \ell^- \ell^{\prime +}$.

It turns out that
\begin{align}
\mathcal{B}(B \to K \tau\mu) \approx 4 \times 10^{-8} \left| C_{9}^{23}\right|^2 
\approx 10^{-7} 
\left| \frac{C_{9}^{22}}{0.5}\, \frac{0.3}{\lambda^e_{23}}\right|^2,
\end{align}
where we have used the relation $C_{9}^{22}/C_{9}^{23} \!\approx \! \lambda^e_{23}$ and we recall that, 
in order to accommodate the $R^{e/\mu}_K$ anomaly, $|C_{9}^{22}| \approx 0.5$.
The above prediction is far below the current experimental bound $\mathcal{B}(B\to K \tau\mu) \leq 4.8 \times 10^{-5}$~\cite{Amhis:2014hma}.
Moreover, we find
\begin{align}
\mathcal{B}(B\to \tau^\pm \mu^\mp) &\approx \mathcal{B}(B\to K\tau^\pm \mu^\mp)\,,
\\
\mathcal{B}(B\to K^*\tau^\pm \mu^\mp) &\approx 2 \times \mathcal{B}(B\to K \tau^\pm \mu^\mp)
\,.
\end{align}
%
As we will see shortly, loop-induced $\tau$ LFV decays are typically better probes of our scenario than LFV $B$ decays.

\subsection{One-loop induced LFV and LFUV phenomenology}
As illustrated in Section 2, electroweak corrections give rise to modified couplings of the $Z$ and $W$ bosons to leptons and to 
a purely leptonic low-energy effective Lagrangian. LFV and LFUV are both expected in $Z$, $W$ and $\tau$ lepton decays. 
In this section we analyze these processes providing approximate analytical expressions for the corresponding observables in our framework.

\subsubsection{\texorpdfstring{$Z \to \ell \ell^\prime$ and $Z \to \nu \nu^\prime$}{Z to ll, nn}}
%
At the loop-level the leptonic $Z$ couplings are modified in our framework. 
Their departure from the SM expectations are constrained by the LEP measurements
of the $Z$ decay widths, left-right and forward-backward asymmetries.
The bounds on lepton non-universal couplings are reported in table~\ref{tab:LEP_asym}
and are expressed in terms of the vector and axial-vector couplings $v_\ell$ and $a_\ell$, 
respectively, defined as 
\begin{equation}
v_\ell = g^{\ell\ell}_{\ell L} + g^{\ell\ell}_{\ell R} \qquad\qquad
a_\ell = g^{\ell\ell}_{\ell L} - g^{\ell\ell}_{\ell R} 
\,.
\end{equation}
We get
\begin{equation}
\frac{v_\tau}{v_e} = 1 - \frac{2\,\delta g^{33}_{\ell L}}{(1-4s^2_W)}\qquad\qquad
\frac{a_\tau}{a_e} = 1 - 2\,\delta g^{33}_{\ell L}
\,,
\end{equation}
with $\delta g^{ij}_{\ell L}$ defined in eq.~(\ref{newgeL}), leading to the following numerical estimates
\begin{align}
\frac{v_\tau}{v_e} & \approx 1 - 0.05 \, \frac{\left( C_1 - 0.8 \,C_3 \right)}{\Lambda^2({\rm TeV^2})}\,, 
\\
\frac{a_\tau}{a_e} & \approx 1 - 0.004 \, \frac{\left( C_1 - 0.8 \,C_3 \right)}{\Lambda^2({\rm TeV^2})}\,,
\label{eq:zpole_num}
\end{align}
where we took $\lambda^u_{33}\sim \lambda^e_{33}\simeq 1$ and, hereafter, we set $\Lambda = 1$ TeV in the argument of the logarithm.
%
\begin{table}[tb]
\centering
\vspace{0.2cm}
\renewcommand{\tabcolsep}{0.7pc} 
\renewcommand{\arraystretch}{1.2} 
\begin{tabular}{|c|ccc|}
\hline 
& $e$ & $\mu$ & $\tau$ 
\\ 
\hline &&&\\[-8mm]
$v_\ell$ & $-0.03817\; (47)$ & $-0.0367\; (23)$ & $-0.0366\; (10)$ 
\\
$a_\ell$ & $-0.50111\; (35)$ & $-0.50120\; (54)$ & $-0.50204\; (64)$
\\
[2mm] \hline\hline \multicolumn{4}{|c|}{}\\[-8mm]
\multicolumn{4}{|c|}{$v_\mu/v_e = 0.961\; (61) \qquad\qquad\qquad a_\mu/a_e = 1.0002\; (13)$}
\\
\multicolumn{4}{|c|}{$v_\tau/v_e = 0.959\; (29) \qquad\qquad\qquad a_\tau/a_e = 1.0019\; (15)$}
\\[2mm] \hline
\end{tabular}
\caption{Vector and axial-vector $Z$ couplings from the measured values of the leptonic $Z$ decay widths, 
left-right and forward-backward asymmetries of the final lepton $\ell^-$ (from PDG~\cite{Agashe:2014kda}). 
\label{tab:LEP_asym}}
\end{table}
%
%
Moreover, modifications of the $Z$ couplings to neutrinos affect the extraction of the number 
of neutrinos $N_\nu$ from the invisible Z decay width. We find that
\begin{align}
N_\nu = 2 + \, \left(\frac{g^{33}_{\nu L}}{g^{\mysmall \rm SM}_{\nu L}}\right)^2 \simeq\, 3 + 4 \, \delta g^{33}_{\nu L}\,,
\end{align}
with $\delta g^{ij}_{\nu L}$ defined in eq.~(\ref{newgnuL}),
leading to the following numerical estimate
\begin{equation}
N_\nu \approx 3 + 0.008 \, \frac{\left( C_1 + 0.8 \,C_3 \right)}{\Lambda^2({\rm TeV^2})}\,,
\end{equation}
to be compared with the experimental result~\cite{Agashe:2014kda}
\begin{equation}
N_\nu = 2.9840 \pm 0.0082\,.
\end{equation}
Finally, we consider the LFV decay modes of the Z boson, $Z \to \ell^\pm_f \ell^\mp_i$, described by the following branching ratio
\begin{equation}
\mathcal{B}(Z \to \ell^\pm_i \ell^\mp_j) \simeq \frac{G_{\rm F} \, m^3_Z}{3\pi \sqrt{2} \, \Gamma_Z} 
\left[ 
(g^{ij}_{\ell L} + \delta g^{ij}_{\ell L} )^2 + (g^{ij}_{\ell R} )^2 - \frac{3}{4}\frac{m^2_{\ell_i}}{m^2_Z} \, \delta_{ij}
\right]\,,
\label{eq:Zllprime}
\end{equation}
where $\Gamma_Z \approx 2.5 $ GeV and therefore we obtain
\begin{equation}
\mathcal{B}(Z \to \mu^\pm \tau^\mp) \approx
10^{-7}~\frac{\left[\lambda^{u}_{33}(C_1-C_3) + 0.25 C_3\right]^2}{\Lambda^4({\rm TeV^4})}
\left(
\frac{\lambda^e_{23}}{0.3}
\right)^2
\,,
\end{equation}
which is well below the current experimental bound $\mathcal{B}(Z \to \mu^\pm \tau^\mp)_{\rm exp} \leq 1.2 \times 10^{-5}$,
especially when $C_1=C_3$.

\subsubsection{\texorpdfstring{$W \to \ell \nu$}{W --> l nu}}
%
At the loop-level also the leptonic $W^\pm$ couplings $g^{ij}_{\ell}$ are modified with respect to their SM expectations. 
In particular, summing the RGE contributions, see eq.~(\ref{deltas}), and the relevant matrix element, the NP corrections 
to $g^{ij}_{\ell}$ in our model read
\begin{equation}
\delta g_{\ell}^{ij}=\frac{v^2}{\Lambda^2} \, \frac{\log(\Lambda/m_{\mysmall EW})}{16\pi^2}\,
(-2 g_2^2 C_3+6 y_t^2 \lambda^u_{33} C_3)\lambda^{e}_{ij}~,
\end{equation}
up to finite corrections.
Let us define now the quantity $R^{\tau/\ell}_{W}$ which accounts for LFUV in $W^\pm$ decays 
\begin{align}
R^{\tau/\ell}_{W} = 
\frac{\mathcal{B}(W \to \tau\nu)_{\rm exp}/\mathcal{B}(W \to \tau\nu)_{\rm SM}}{\mathcal{B}(W \to \ell\nu)_{\rm exp}/\mathcal{B}(W \to \ell\nu)_{\rm SM}} \,,\qquad\qquad \ell=e,\mu~.
\end{align}
In our setup, $R^{\tau/\ell}_{W}$ is given by
\begin{equation}
R^{\tau/\ell}_{W}  \approx 1 + 2\, \delta g^{33}_{\ell}
\approx 1 + \frac{0.008 \, C_3}{\Lambda^2({\rm TeV^2})}\,,
\label{eq:Wln}
\end{equation}
which has to be compared with the LEP measurements~\cite{Agashe:2014kda}
\begin{equation}
R^{\tau/\mu}_{W}  = 1.068(26)\,,\qquad R^{\tau/e}_{W} = 1.062(26)\,.
\end{equation}
As we will see shortly, LFUV in $\tau$ decays will provide more stringent constraints
on our model parameters than $W^\pm$ decays.

\subsubsection{\texorpdfstring{$\tau\to \ell \nu\bar\nu$}{tau --> l nu nu}}
\label{sec_tauLFU}

The effective Lagrangian relevant for charged-current processes like $\tau\to \ell \nu\bar\nu$ is given by
\begin{equation}
{\cal{L}}^{\mysmall \rm CC}_{\rm eff} \!=\! -\frac{4 G_F}{\sqrt{2}} \, (C^{\tau\ell}_L)_{ij}
\left( \bar \nu_{iL} \gamma_\mu \tau_L  \right) \left( \bar\ell_L \gamma^\mu \nu_{jL}  \right) + h.c.\,,
\end{equation}
where $\ell =e,\mu$ and the coefficients $(C^{\tau\ell}_L)_{ij}$ read
\begin{equation}
(C^{\tau\ell}_L)_{ij} = \delta_{i3}\delta_{\ell j} + \frac{v^2}{\Lambda^2} y^2_t \lambda^{u}_{33} 
\left[
3 (C_1-C_3) \delta_{ij} \lambda^{e}_{\ell 3} 
+ 
6 C_3 
(
\delta_{\ell j} \lambda^{e}_{i3} + \delta_{i3} \lambda^{e}_{\ell j}
)
\right] \frac{\log(\Lambda/m_{EW})}{16\pi^2}
\,.
\end{equation}
Notice that the term in $(C^{\tau\ell}_L)_{ij}$ proportional to $\delta_{ij} \lambda^{e}_{\ell 3} $ generates exclusively 
LFV processes, while the one proportional to $ \delta_{i3} \lambda^{e}_{\ell j} + \delta_{\ell j} \lambda^{e}_{i3}$
contains also lepton flavor conserving contributions which can therefore interfere with the SM amplitude.
We stress that the NP part of $(C^{\tau\ell}_L)_{ij}$ is entirely generated by running effects from $\Lambda$ to $m_{EW}$ driven by
the top yukawa interactions (and therefore proportional to $y^2_{t}$) while gauge contributions (proportional to $g^2_{1,2}$) are absent. 
Moreover, the operator $\left( \bar \nu_{iL} \gamma_\mu \tau_L  \right) \left( \bar\ell_L \gamma^\mu \nu_{jL}  \right)$ is not 
renormalised below the EW scale by QED interactions. The latter point can be easily understood using the Fiertz identity 
$\left( \bar \nu_{iL} \gamma_\mu \tau_L  \right) \left( \bar\ell_L \gamma^\mu \nu_{jL}  \right)=
\left( \bar \nu_{iL} \gamma_\mu \nu_{jL}  \right) \left( \bar\ell_L \gamma^\mu \tau_L\right)$
and remembering that the charged lepton current $\left( \bar\ell_L \gamma^\mu \tau_L\right)$
is protected from renormalization effects by the QED Ward identity which stems from 
the electric charge conservation. LFUV in $\tau \to \ell \nu\bar\nu$ is described by the observables
\begin{align}
R^{\tau/e}_\tau = \frac{\mathcal{B}(\tau \!\to\! \mu \nu\bar\nu)_{\rm exp}/\mathcal{B}(\tau \!\to\! \mu \nu\bar\nu)_{\rm SM}}{\mathcal{B}(\mu \!\to\! e \nu\bar\nu)_{\rm exp}/\mathcal{B}(\mu \!\to\! e \nu\bar\nu)_{\rm SM}} \qquad
R^{\tau/\mu}_\tau = \frac{\mathcal{B}(\tau \!\to\! e \nu\bar\nu)_{\rm exp}/\mathcal{B}(\tau \!\to\! e \nu\bar\nu)_{\rm SM}}{\mathcal{B}(\mu \!\to\! e \nu\bar\nu)_{\rm exp}/\mathcal{B}(\mu \!\to\! e \nu\bar\nu)_{\rm SM}} \,,
\end{align}
and are experimentally tested at the few \textperthousand~level~\cite{Pich:2013lsa}
\begin{align}
R^{\tau/\mu}_\tau = 1.0022 \pm 0.0030 \qquad\qquad R^{\tau/e}_\tau = 1.0060 \pm 0.0030 \,.
\label{eq:tau_LFU_data}
\end{align}
$R^{\tau/\ell}_\tau$ can be expressed in terms of the coefficients $(C^{\tau\ell}_L)_{ij}$ as follow,
\begin{equation}
R^{\tau/e}_{\tau} = \frac{\sum_{ij} |(C^{\tau\mu}_{L})_{ij}|^2}{\sum_{ij} |(C^{\mu e}_{L})_{ij}|^2}
\qquad\qquad
R^{\tau/\mu}_{\tau} = \frac{\sum_{ij} |(C^{\tau e}_{L})_{ij}|^2}{\sum_{ij} |(C^{\mu e}_{L})_{ij}|^2}\,.
\label{eq:R_D}
\end{equation}
Keeping only linear terms in the NP contributions we find
\begin{align}
R^{\tau/\ell}_\tau &\simeq 1 + \frac{v^2}{\Lambda^2} y^2_t \lambda^{u}_{33} \lambda^{e}_{33}\,
\frac{3C_3}{4\pi^2}\log \frac{\Lambda}{m_{EW}} 
\approx 1 + \frac{0.008 \, C_3}{\Lambda^2({\rm TeV^2})}
\label{eq:tau_LFU}
\,,
\end{align}
where in the last approximation we have set $y_t =\lambda^{u}_{33} =\lambda^{e}_{33} \simeq 1$ 
and $\Lambda=1$ TeV in the logarithm.
%
\subsubsection{Neutrino trident production}

Our purely leptonic effective Lagrangian induces also the neutrino-nucleus scattering $\nu_\mu N \to \nu_\mu N \mu^+ \mu^-$,
the so-called neutrino trident production (NTP) process (see, e.g. Ref.~\cite{Altmannshofer:2014pba}). Within the SM, NTP 
is generated by the exchange of both the $W$ and $Z$ vector bosons. The effective Lagrangian relevant for the NTP is 
\begin{align}
 &
 \mathcal L^{\rm NTP}_\text{eff}  
 = -\frac{G_F}{\sqrt{2}} \left[ \bar\mu \gamma^\mu \left( C_V - C_A \gamma^5 \right) \mu \right] \left[ \bar\nu \gamma_\mu (1-\gamma^5) \nu \right]\,,  
 \label{EQ:effectiveop}
\end{align}
where $C_{V/A} = C^{\rm SM}_{V/A} + C^{\rm NP}_{V/A}$, $C^{\rm SM}_V = \frac{1}{2} + 2 s^2_\theta$ and $C^{\rm SM}_A = \frac{1}{2}$.
In our model, $C^{\rm NP}_{V/A}$ read 
\bea
 C^{\rm NP}_V &=& - \frac{v^2}{32\pi^2 \Lambda^2}
 \left[ 
 \left(\xi^{\nu e_L}_{2222} + \xi^{\nu e_R}_{2222}\right) \log\frac{\Lambda}{m_{\mysmall EW}} 
 +
2 \, \delta \xi^{\nu e}_{2222} \log\frac{m_{\mysmall EW}}{\mu} 
 \right] 
  \,,\\
 C^{\rm NP}_A &=& - \frac{v^2}{32\pi^2 \Lambda^2}
 \left[ 
 \left(\xi^{\nu e_L}_{2222} - \xi^{\nu e_R}_{2222}\right) \log\frac{\Lambda}{m_{\mysmall EW}} 
 \right] 
  \,,
\eea
where 
\begin{align}
 \xi^{\nu e_L}_{2222} &\!=\!   
\lambda^e_{22} \lambda^u_{33} y^2_t \left[ 12 C_3 -12 s_\theta^2 (C_1 + C_3) \right]
+\frac{4}{3}e^2 \lambda^e_{22} \left(C_1 + 3C_3\right)
  \,,\\
 \xi^{\nu e_R}_{2222} &\!=\!  -12 \lambda^e_{22} \lambda^u_{33} y^2_t s_\theta^2 \left( C_1  + C_3 \right)
 +\frac{4}{3}e^2 \lambda^e_{22} \left(C_1 + 3C_3\right)
  \,,\\
 \delta\xi^{\nu e}_{2222} &\!\!=\! \lambda^e_{22} \frac43 e^2 \! \left[\!(C_1\!+\!3C_3) 
 \!- 2(C_1\!+\!C_3) \!\left(\! \hat{\lambda}^u_{33}\log\frac{m_t}{\mu} \!+\! \hat{\lambda}^u_{22}\log\frac{m_c}{\mu} \!\right) 
 \!\!+\! (C_1\!-\!C_3)\hat{\lambda}^d_{33}\log\frac{m_b}{\mu}  \right]\!.
\end{align}
In terms of the coefficients $C_V$ and $C_A$, the inclusive cross section is proportional to 
$C_V^2 + C_A^2$~\cite{Altmannshofer:2014pba} and therefore, at leading order in the NP contribution, we have
\begin{align}
\left( \frac{ \sigma_\text{SM+NP}}{\sigma_\text{SM}}\right)_{\rm NTP}   
&\simeq  1 + 2~ \frac{(C^{\rm SM}_V C^{\rm NP}_V + C^{\rm SM}_A C^{\rm NP}_A)}{(C^{\rm SM}_V)^2 + (C^{\rm SM}_A)^2}\,,
\end{align}
with the running scale $\mu$ replaced by a constituent mass $\hat\mu\approx 1$ GeV.
On the other hand, the experimental data and the SM prediction are in the following ratio~\cite{Altmannshofer:2014pba}
\begin{align} 
 \left( \frac{\sigma_\text{exp.}}{\sigma_\text{SM}} \right)_{\rm NTP} = 0.82 \pm 0.28 \,. 
\end{align}
From a numerical analysis we find that 
$({ \sigma_\text{SM+NP} / \sigma_\text{SM} })_{\rm NTP}-1 \approx -8 \times 10^{-3} (C_3-0.5C_1)\lambda^e_{22}$
quite below the current experimental resolution.

\subsubsection{\texorpdfstring{$\tau$}{tau} LFV}
%
The purely leptonic and semileptonic parts of our effective Lagrangian generate LFV processes such as $\tau \!\to\! \mu\ell\ell$ and 
$\tau \!\to\! \mu P$ with $P \!=\! \pi,\eta,\eta^\prime,\rho$, etc. In the case of $\tau\to \mu\ell\ell$ we find
\begin{align}
\frac{\mathcal{B}(\tau\to \mu\ell\ell)}{\mathcal{B}(\tau\to \mu\nu\bar\nu)} 
= 
\left(\frac{v^2}{32\pi^2 \Lambda^2}\right)^2
\bigg[
(1+\delta_{\ell\mu})
&\left(
\xi^{e_L e_L}_{23\ell\ell} \log\frac{\Lambda}{m_{\mysmall EW}}  + \delta\xi^{e_L e}_{23\ell\ell} \log\frac{m_{\mysmall EW}}{\hat\mu} 
\right)^2
+
\nn\\
&\left(
\xi^{e_L e_R}_{23\ell\ell} \log\frac{\Lambda}{m_{\mysmall EW}}  + \delta\xi^{e_L e}_{23\ell\ell} \log\frac{m_{\mysmall EW}}{\hat\mu} 
\right)^2
\bigg]~,
\end{align}
where
\begin{align*}
\xi^{e_L e_L}_{23\ell\ell} &=  \lambda^e_{23} 
	\left[  \frac43 e^2(C_1-3C_3)-12(-\frac12+s^2_\theta) \, y_t^2 \lambda^u_{33}(C_1-C_3)  \right]
	\\
\xi^{e_L e_R}_{23\ell\ell} &=  	\lambda^e_{23} 
								\left[  \frac43 e^2(C_1-3C_3)-12 \, s^2_\theta \, y_t^2 \lambda^u_{33}(C_1-C_3)  \right]
	\\
\delta\xi^{e_L e}_{23\ell\ell} &= 	\lambda^e_{23} \, \frac43 e^2
\left[  (C_1-3C_3) - 2(C_1-C_3) (\hat{\lambda}^u_{33}\log\frac{m_t}{\hat\mu}+\hat{\lambda}^u_{22}\log\frac{m_c}{\hat\mu})
								 +(C_1+C_3)\hat{\lambda}^d_{33}\log\frac{m_b}{\hat\mu}  \right]~.
\end{align*}
where $\hat\mu\approx 1$ GeV is the common constituent mass for the light quarks.
If $C_1-C_3 \sim \mathcal{O}(1)$, the leading effects to $\mathcal{B}(\tau\to \mu\ell\ell)$ stem from top-yukawa interactions 
and we end up with the following numerical estimate
\begin{equation}
\mathcal{B}(\tau\to 3\mu) \approx 5\times 10^{-8} \,
\frac{(C_1-C_3)^{\,2} }{\Lambda^4({\rm TeV^{4}})} 
\left(\frac{\lambda^e_{23}}{0.3}\right)^2 \,,
\label{eq:tau3mu_num}
\end{equation}
to be compared with the current experimental bound $\mathcal{B}(\tau\to 3\mu) \leq 1.2 \times 10^{-8}$~\cite{Amhis:2014hma}.
Notice that in our framework it turns out that $1.5 \lesssim \mathcal{B}(\tau \!\to\! 3\mu)/\mathcal{B}(\tau \!\to\! \mu ee)\lesssim 2$.
On the other hand, in scenarios where $C_1=C_3$, top-yukawa contributions vanish and $\mathcal{B}(\tau\to 3\mu)$
is dominated by the electromagnetic interactions. The resulting $\mathcal{B}(\tau\to 3\mu)$ for $\Lambda = 1$ TeV, 
$\lambda^e_{23}=0.3$ and $C_1=C_3=1$ is $\mathcal{B}(\tau\to 3\mu) \approx 4\times 10^{-9}$, yet within the future 
expected experimental sensitivity.

We turn now to the processes $\tau\to \mu\rho$ and $\tau\to \mu\pi$. Employing the general formulae of Ref.~\cite{Brignole:2004ah}, 
we find
\begin{align}
\frac{\mathcal{B}(\tau\to \mu\rho)}{\mathcal{B}(\tau\to \nu_\tau\rho)} 
= 
\left(\frac{v^2}{32\pi^2 \Lambda^2}\right)^2\frac{1}{2\cos^2\theta_c}
\bigg(
&(\xi^{e_L u_L}_{2311} - \xi^{e_L d_L}_{2311} + \xi^{e_L u_R}_{2311} - \xi^{e_L d_R}_{2311}) \log\frac{\Lambda}{m_{\mysmall EW}}  +
\nn\\
&2(\delta\xi^{e_L u}_{2311} - \delta\xi^{e_L d}_{2311}) \log\frac{m_{\mysmall EW}}{\hat\mu} 
\bigg)^2\,,
\end{align}
where $\theta_c$ is the Cabibbo angle, $\mathcal{B}(\tau\to \nu_\tau\rho)\approx 25\%$ and we have defined
\begin{align*}
\xi^{e_L u_L}_{2311} &= \lambda^e_{23}\left[  -\frac89 e^2(C_1-3C_3)-12(\frac12-\frac32 s^2_\theta) y_t^2 \lambda^u_{33}(C_1-C_3)  \right]~,
\\
\xi^{e_L d_L}_{2311} &= \lambda^e_{23} \left[  \frac49 e^2(C_1-3C_3)-12(-\frac12+\frac13 s^2_\theta)\, y_t^2 \lambda^u_{33}(C_1-C_3)  \right]~,
\\
\xi^{e_L u_R}_{2311} &= \lambda^e_{23}\left[   -\frac89 e^2(C_1-3C_3)+8 \, s^2_\theta \, y_t^2 \lambda^u_{33}(C_1-C_3)  \right]~,
\\
\xi^{e_L d_R}_{2311} &= \lambda^e_{23}	\left[  \frac49 e^2(C_1-3C_3)-4 \, s^2_\theta\, y_t^2 \lambda^u_{33}(C_1-C_3)  \right]~,
\\
\delta\xi^{e_L u}_{2311} &= - \lambda^e_{23} \cdot
	     		 \frac89 e^2\left[(C_1-3C_3)-2(C_1-C_3)(\hat{\lambda}^u_{33}\log\frac{m_t}{\hat\mu}	+\hat{\lambda}^u_{22}\log\frac{m_c}{\hat\mu})\right]~,
\\
\delta\xi^{e_L d}_{2311} &= \lambda^e_{23} \, \frac49 e^2
								\left[(C_1-3C_3)-2(C_1-C_3)(\hat{\lambda}^u_{33}\log\frac{m_t}{\hat\mu}
								+\hat{\lambda}^u_{22}\log\frac{m_c}{\hat\mu})+(C_1+C_3)\hat{\lambda}^d_{33}\log\frac{m_b}{\hat\mu}\right] \,.
\end{align*}
A numerical estimate for $\mathcal{B}(\tau\to\mu\rho)$ is given by
\begin{align}
\mathcal{B}(\tau\to\mu\rho) \approx 5 \times 10^{-8} \frac{(C_1 - 1.3 \,C_3)^2}{\Lambda^4({\rm TeV^{4}})} 
\left(\frac{\lambda^e_{23}}{0.3}\right)^2
\,,
\end{align}
where the current bound is $\mathcal{B}(\tau\to \!\mu\rho) \leq 1.5 \times 10^{-8}$.
Finally, the $\mathcal{B}(\tau\to \mu\pi)$ expression is
\begin{align}
\frac{\mathcal{B}(\tau\to \mu\pi)}{\mathcal{B}(\tau\to \nu_\tau\pi)} 
= 
\left(\frac{v^2}{32\pi^2 \Lambda^2}\right)^2\frac{1}{2\cos^2\theta_c}
\bigg(
&(\xi^{e_L u_L}_{2311} - \xi^{e_L d_L}_{2311} - \xi^{e_L u_R}_{2311} + \xi^{e_L d_R}_{2311}) \log\frac{\Lambda}{m_{\mysmall EW}} 
\bigg)^2\,,
\end{align}
where $\mathcal{B}(\tau\to \nu_\tau\pi)\approx 11\%$. We notice that, since the $\pi$ meson 
is a pseudoscalar, $\mathcal{B}(\tau\to \mu\pi)$ does not receive contributions from electromagnetic interactions. The full result is 
well approximated by the following numerical expression
\begin{align}
\mathcal{B}(\tau\to\mu\pi) 
\approx 8 \times 10^{-8} \, \frac{(C_1-C_3)^2}{\Lambda^4({\rm TeV^{4}})} 
\left(\frac{\lambda^e_{23}}{0.3}\right)^2
\,,
\label{eq:taumurho_num}
\end{align}
where the current bound reads $\mathcal{B}(\tau\to\mu\pi) \leq 2.7\times 10^{-8}$~\cite{Amhis:2014hma}.

\begin{figure}[p]
\centering
\begin{minipage}{0.45\textwidth}
\includegraphics[width=\textwidth]{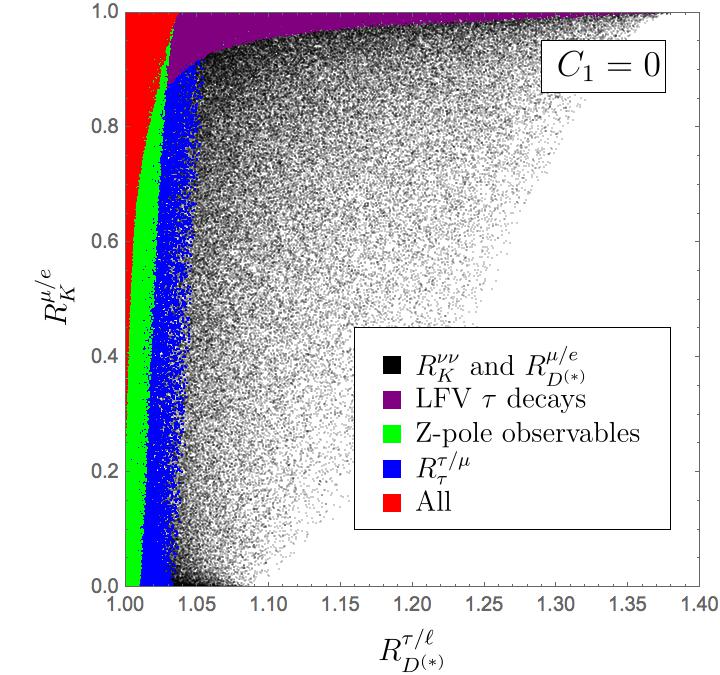}
\end{minipage}
\begin{minipage}{0.45\textwidth}
\includegraphics[width=\textwidth]{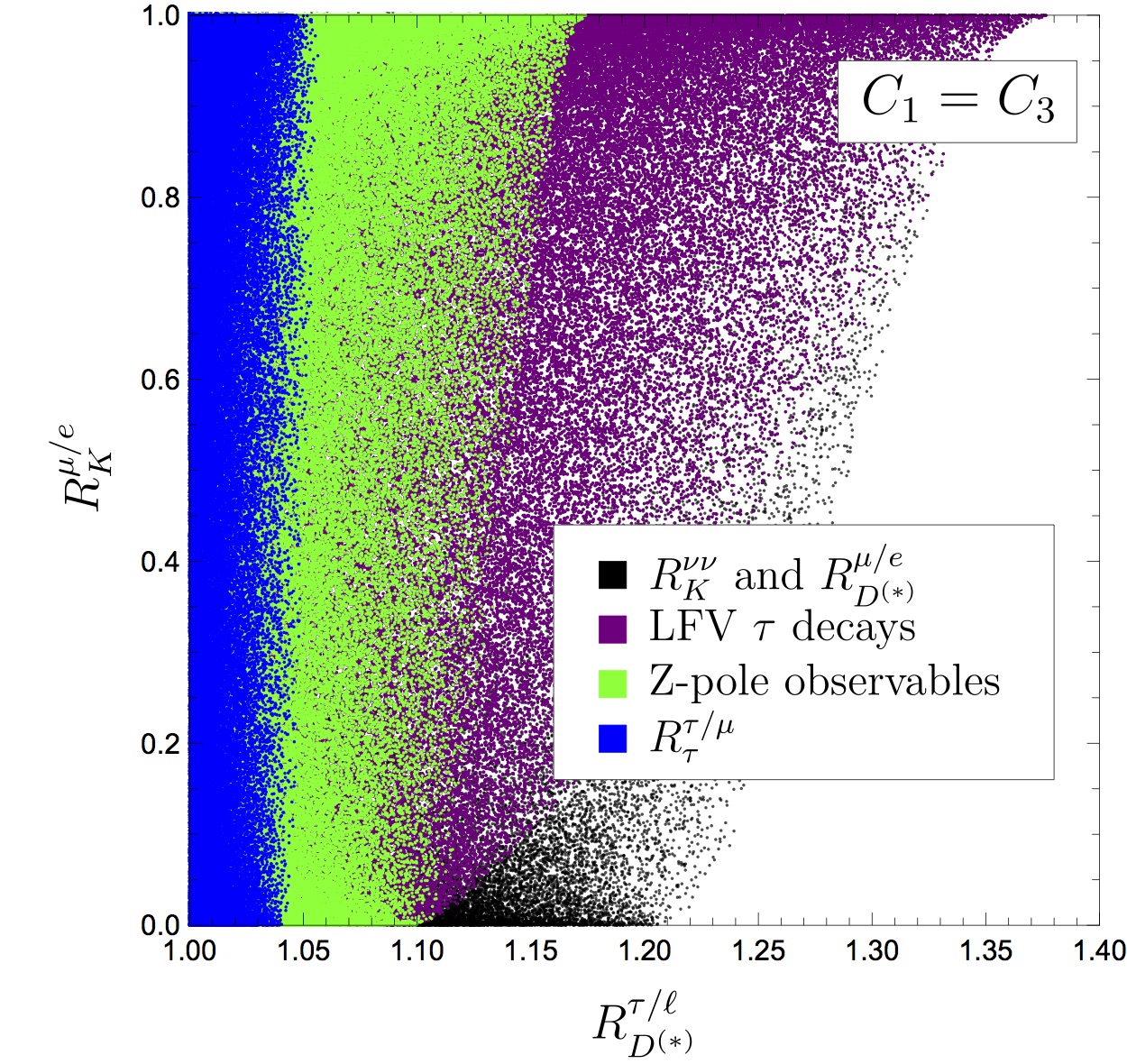}
\end{minipage}
\caption{Impact of one-loop-triggered constraints when addressing the $B$ anomalies through left-handed currents, for two different $C_{1}$ vs.~$C_3$ configurations (\emph{left}: $C_1=0$, \emph{right}: $C_1 = C_3$).
For $C_1=C_3$, simultaneously imposing all bounds is actually equivalent to impose $R_\tau^{\tau/\ell}$ alone.
In the scan the parameters varied in the following ranges: $C_{1,3}/\Lambda^2  \in \{-4,4\}~{\rm TeV}^{-2}$, $\Lambda \in \{1,10\} \TeV$, $ |\lambda^{d,e}_{23} | \in \{0,0.5\}$. All bounds refer to $2\sigma$ uncertainties.
				\label{fig_numerical1}}
\end{figure}

\begin{figure}[p]
\centering
\begin{minipage}[t]{0.45\textwidth}
\includegraphics[width=\textwidth]{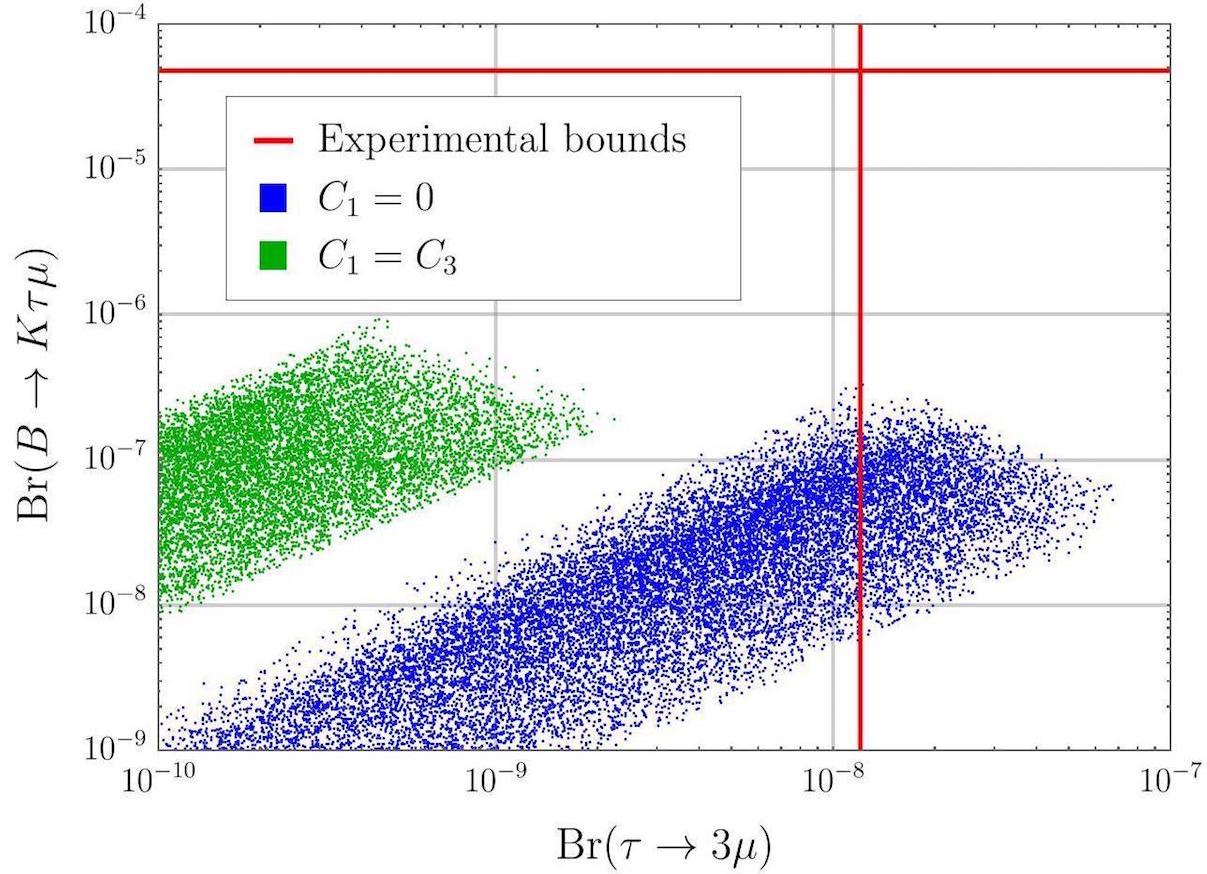}
\end{minipage}
~~~
\begin{minipage}[t]{0.45\textwidth}
\includegraphics[width=\textwidth]{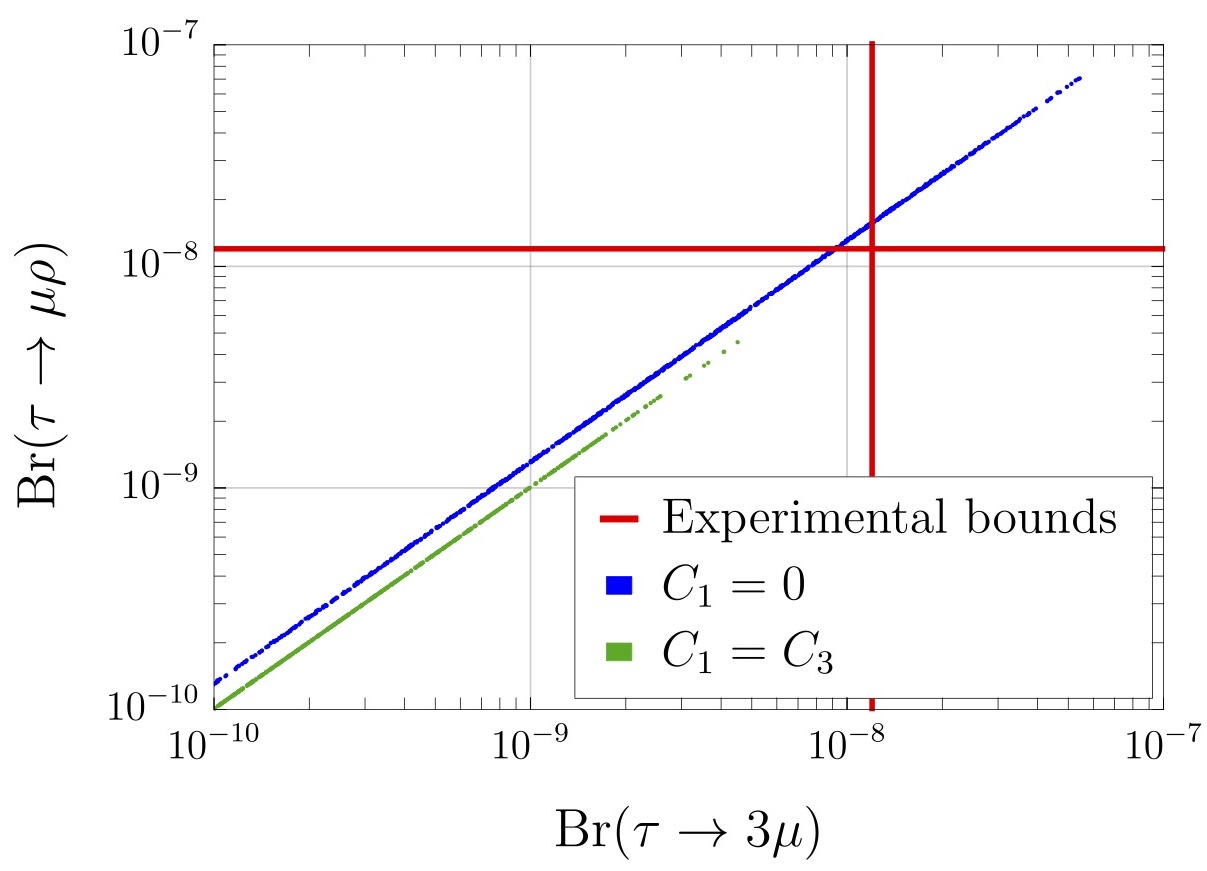}
\end{minipage}
\caption{
		{\it Left (right)}: Correlation ${\rm Br}(\tau \to 3 \mu)$ vs.~${\rm Br}(B \to K \tau \mu)$ (${\rm Br}(\tau \to 3 \mu)$ vs.~${\rm Br}(\tau \to \mu \rho)$) 
		within our model, while satisfying all other bounds but $R_{D^{(*)}}^{\tau / \ell}$, for two different $C_{1}$ vs.~$C_3$ configurations.
		In the scan the parameters varied in the following ranges: $C_{1,3}/\Lambda^2  \in \{-4,4\}~{\rm TeV}^{-2}$, $\Lambda \in \{1,10\} \TeV$, $ |\lambda^{e}_{23} | \in \{0,0.5\}$, $ \lambda^{d}_{23}  \in \{-0.2, -0.01\}$. All bounds refer to $2\sigma$ uncertainties.
				\label{fig_numerical2}}
\end{figure}

\begin{figure}[tp]
\centering
\includegraphics[width=0.5\textwidth]{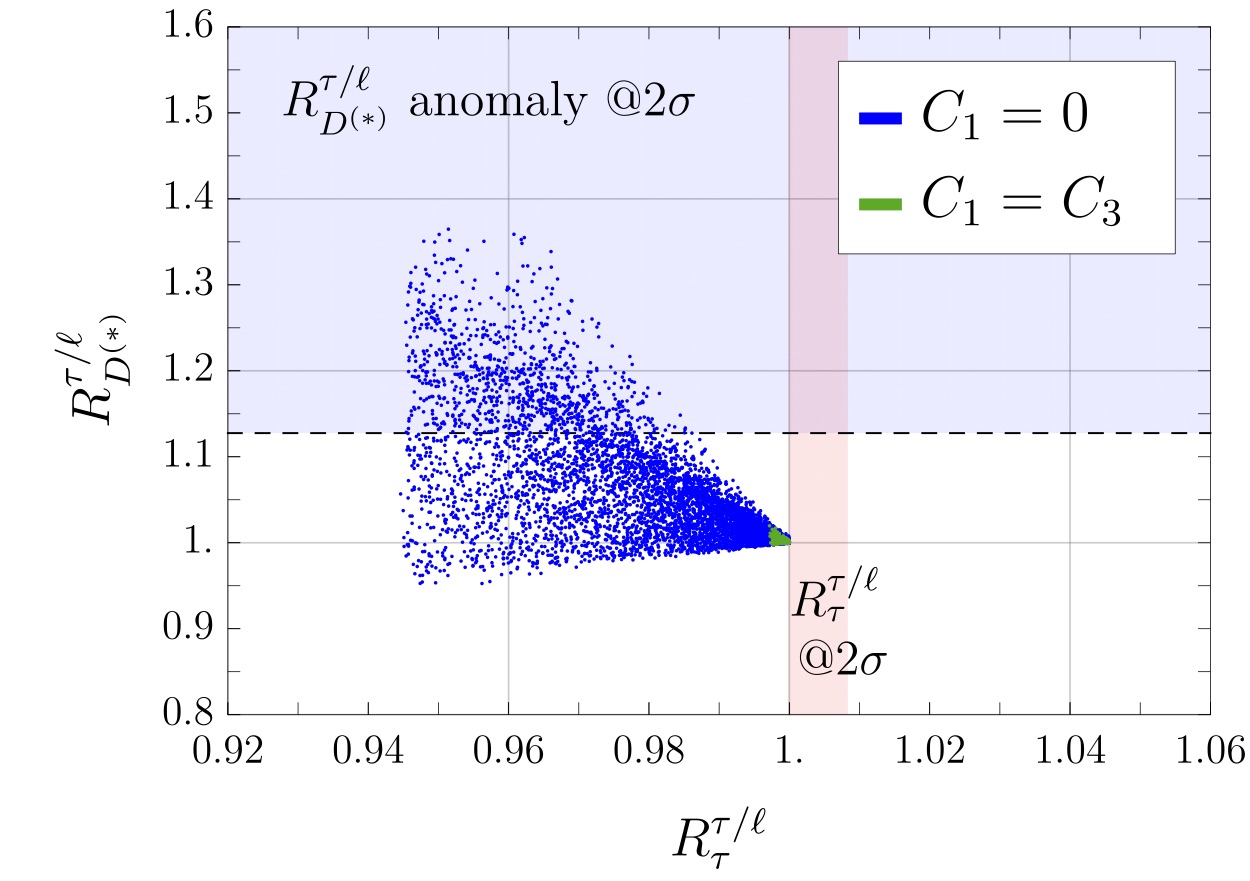}
\caption{ 
		Correlation between $R_\tau^{\tau/\ell}$ and $R_{D^{(*)}}^{\tau / \ell}$ predictions when scanning the parameter space of the model
	    (see Fig.~\ref{fig_numerical2} for details about the scan). The $2\sigma$ lower limit for the $R_{D^{(*)}}^{\tau / \ell}$ anomaly and the
	     combined $2\sigma$ bounds of $R_\tau^{\tau/\mu}$ and $R_\tau^{\tau/e}$ are also shown.
				\label{fig_numerical3}}
\end{figure}

\subsection{Numerical analysis}
In previous paragraphs we have analysed the challenges raised by individual observables to a common  explanation of $B$ anomalies 
through left-handed currents, the main outcome being that LFV and LFUV processes triggered at one loop play a major role in constraining 
such scenarios.
However, it is only through a global analysis that we can appreciate the interplay between different constraints and correctly quantify the 
effects induced by the full low-energy effective Lagrangian derived in this paper.
As discussed at the beginning of this section, we can fully parametrize our setup using $C_{1,3}/\Lambda^2$, $\lambda^{d,e}_{23}$. 
It is thus relatively easy to scan the parameter space of the model. We let the parameters vary within the following windows:
\bea
\label{param}
-4~ {\rm TeV}^{-2} \le& \displaystyle\frac{C_{1,3}}{\Lambda^2}\le& +4~ {\rm TeV}^{-2}~~~~~~~1~ {\rm TeV}\le \Lambda\le 10~ {\rm TeV} \nonumber\\
0\le& |\lambda^{e}_{23} |\le& 0.5\\
0\le& |\lambda^{d}_{23} |\le& 0.5~~~~~~~{\rm or}~~~~~~~0.01\le |\lambda^{d}_{23} |\le 0.2~.
\nonumber
\eea
We display our main results in fig.~\ref{fig_numerical1}, \ref{fig_numerical2} and \ref{fig_numerical3}, where the obtained limits are 
defined by the $2\sigma$ uncertainties of the observable quantities, assumed to be uncorrelated. Notice that the choice 
$C_{1,3}/\Lambda^2  \in \{-4,4\}~{\rm TeV}^{-2}$ is by no means restrictive, since values out of this range are excluded by $Z$ decays 
and $\tau$ LFUV phenomenology.
Moreover, although the full allowed range for $ |\lambda^{d}_{23} |$ is $\{0,0.5\}$, it is reasonable to assume $| \lambda^{d}_{23} |$ 
to be close to the $V_{cb}$ value, in order to avoid too much fine tuning when reproducing the $V_{CKM}$ matrix. This is the reason 
for the chosen $| \lambda^{d}_{23} |$ range in Figs.~\ref{fig_numerical2}, \ref{fig_numerical3}, the sign being the correct one to 
reproduce the $\RKss$ anomaly.

In fig.~\ref{fig_numerical1} we can appreciate the impressive impact of the LFV and LFUV bounds from lepton phenomenology, 
for two typical choices of the parameters $C_{1,3}$, namely $C_1=0$ and $C_1=C_3$.
The black dots are parameter configurations allowed by tree-level semileptonic bounds, i.e. those discussed in section 3.2.
When $C_1=0$, LFUV in $Z$ decays represents the single most powerful constraint, while for $C_1=C_3$ there 
is a partial cancellation of the loop-induced effects in $Z$-pole and LFV observables. In this case the constraint coming from 
$R^{\tau/\mu}_\tau$ is the strongest one. In both cases values of $R_{D^{(*)}}^{\tau / \ell}$ exceeding $1.05$ cannot be accomodated.
In Fig.~\ref{fig_numerical3}, the same conclusion is reinforced by the comparison between the $R_{D^{(*)}}^{\tau / \ell}$ prediction 
and the most challenging LFUV observable, $R^{\tau/\ell}_\tau$, which tests LFUV in purely leptonic decays.
We have collectively called $R_\tau^{\tau/\ell}$ the two observables $R_\tau^{\tau/\mu}$ and $R_\tau^{\tau/e}$ discussed in 
section \ref{sec_tauLFU}, whose predictions are equal at leading order in our model, see eq.~(\ref{eq:tau_LFU}). 
In particular, the experimental value of $R_\tau^{\tau/e}$, which is $2\sigma$ away from the SM prediction and which has not been 
considered in Fig.~\ref{fig_numerical1} to remain on the conservative side, further reduces the allowed $R_{D^{(*)}}^{\tau / \ell}$ 
departure from the SM prediction. Finally, plots of Fig.~\ref{fig_numerical2} deal with the LFV predictions of our model, while satisfying 
all the other bounds, but the $R_{D^{(*)}}^{\tau / \ell}$ anomaly, otherwise no points would survive the scan. Also in these plots we 
choose the two representative values $C_1=0$ and $C_1=C_3$. In the \emph{left} plot, we study the correlation between the predictions 
for the LFV semileptonic $B \to K \tau \mu$ and leptonic $\tau \to 3 \mu$ decays. 
An important result of the present analysis is that the process $\tau \to 3 \mu$ is a much more sensitive probe of the scenario under discussion, 
due both to the better proximity of the predicted ${\rm Br}(\tau \to 3 \mu)$ to the current experimental bound, and to the expected improvements 
of such bound in the near future compared to the challenging semileptonic $B \to K \tau \mu$ process. 
The \emph{right} plot illustrates the comparable potential of other LFV $\tau$ decays compared with $\tau \to 3 \mu$. 
In particular, the $\tau \to \mu \rho$ channel is rather equivalent to $\tau \to 3 \mu$, both in terms of predictions and sensitivity. 
As another example, the $\tau \to \mu \pi$ channel, not displayed in Fig.~\ref{fig_numerical2}, has analogous predictions to 
$\tau \to \mu \rho$ for the $C_1 = 0$ case, but it suffers a cancellation which reduces the predicted branching ratio for 
$C_1 \approx C_3$, as can be seen in eq.~(\ref{eq:taumurho_num}).

In conclusion, a simultaneous explanation of both the $R_{D^{(*)}}^{\tau / \ell}$ and  $\RKss$ anomalies is strongly disfavoured 
in the present framework, where NP at the TeV scale mainly affects left-handed currents\footnote{In principle, such a conclusion 
could be invalidated by large cancellations among different NP contributions entering $R_{\tau}^{\tau / \ell}$ 
with a required fine-tuning of order $10\%$~\cite{Barbieri:2016las,Bordone:2017anc}. So far, no UV-complete model accomplishing 
this task appeared in the literature.}. 
Quite independently on the relative weight between $C_1$ and $C_3$, the current data on $R^{\tau/\ell}_\tau$ forbid values of 
$R_{D^{(*)}}^{\tau / \ell}$ exceeding $1.05$.
Leaving aside the $R_{D^{(*)}}^{\tau / \ell}$ anomaly, in the present framework a significant room is available for LFV effects. 
However, our results contradict the widespread opinion that $B \to K \tau \mu$ is the most promising channel to test these effects,
since the best probes of LFV effects are by far the $\tau$ decays.

\section{Relevance for specific NP models}

In the previous sections we focused on a model-independent analysis of the B anomalies based on the effective Lagrangian 
${\cal L}_{NP}^0(\Lambda)$ of eq. (\ref{LNP}), which was constructed under the following assumptions: 
\begin{itemize}
\item[(i)] NP mainly affects $V-A$ semileptonic fermion currents, 
i.e. only the operators $O_{\ell q}^{(1)}$ and $O_{\ell q}^{(3)}$ are present at the scale $\Lambda$;
\item[(ii)] a basis exists where NP affects only the third fermion generation. Such a basis is approximately aligned to the mass basis in the quark sector. 
\end{itemize}

Notice that we have supplemented the assumption (ii) by the additional requirement of approximate alignment between the quark mass basis
and the basis where NP mainly affects the third generation \footnote{Of course there are two distinct mass basis in the quark sector, slightly 
misaligned by the CKM mixing matrix. The above statement refers to an approximate alignment to any of these two basis.}. Such requirement 
is an output of our analysis and represents a necessary condition to simultaneously accommodate the anomalies in $B\to K\mu^+\mu^-$ and 
$B \to D^{(*)} \tau \nu$.
In our framework this condition translates into the requirement that the unitary matrix $V_d$ is close to the identity matrix. 
In the SM the matrices $V_{u,d}$ are not physical, only their combination $V_{CKM}$ is.
In our setup the NP effects in lighter generations arise through the rotation to the mass basis after EWSB, see eqs.~(\ref{parametrization}) and 
(\ref{lag1}), and are controlled by $V_{u,d}$. In particular we need $|V_{d32}|\ll|V_{d33}|$ to enforce $|\lambda^d_{23}|\ll |\lambda^d_{33}|$
which guarantees a correct interplay between the SM and NP contributions to the NC and CC anomalies, see eqs. (\ref{an1}) and (\ref{an2}).
 
In this section, we want to discuss how far our conclusions depend upon the above assumptions and to what extent such assumptions hold in 
specific NP scenarios. Assumption i) can be realised by the tree-level exchange of a limited number of mediators. 
If we restrict to spin 0 and spin 1 mediators, there are only six possibilities, listed in table 11. 
There are only two color-singlet mediators, either an electroweak singlet $A_\mu$ or a triplet $A^a_\mu$.
Their tree-level exchange gives rise to the operators $O^{(1)}_{lq}$ and $O^{(3)}_{lq}$, respectively.
Therefore, models containing only massive fields $A_\mu$ can address the $\RKss$ but not the $\RD$ 
anomalies~\cite{Gauld:2013qja,Buras:2014fpa,Greljo:2015mma}.
On the other hand, both $A_\mu$ and $A^a_\mu$ generate purely four-lepton and four-quark interactions 
which, in turn, induce very dangerous tree-level effects for processes like $\tau\to 3\mu$ and $B_s-\bar B_s$ mixing. 
In principle, the effects of tree-level four-lepton interactions can compete with or dominate over the loop-induced effects discussed in the present paper~
anomalies~\cite{Gauld:2013qja,Buras:2014fpa,Greljo:2015mma}.

Moreover, there are four types of lepto-quark (LQ) mediators giving rise to $O^{(1)}_{lq}$ and $O^{(3)}_{lq}$ at the tree-level~\cite{Becirevic:2015asa}.
Two have spin one, $U_\mu$ and $U^a_\mu$, and two have spin zero, $S$ and $S^a$. 
In the case of the spin zero electroweak singlet $S$, it turns out that $C_1+C_3=0$ and therefore $\RKss$ is not modified. 
This conclusion remains true even after the inclusion of electroweak RGE effects, at least when NP affects a single generation. 
Out of the LQ mediators, the spin 1 electroweak singlet $U_\mu$ partially evades the bound from $B\to K^{(*)} \nu\bar\nu$, 
since it produces operators satisfying the the tree-level condition $C_1-C_3=0$.
However, as pointed out in Ref.~\cite{Feruglio:2016gvd} and thoroughly discussed in Section 3.2.2, such condition is
not RGE stable and a contribution to $B\to K^{(*)} \nu\bar\nu$ is generated at one-loop via electroweak effects.
Notice that, in contrast to the case of color-singlet mediators, at the tree-level LQs do not give rise neither to four-quark nor to four-lepton operators
and therefore the corresponding indirect constraints from low-energy processes are significantly relaxed.
The new bounds arising from Z-pole observables and $\tau$ decays discussed in Section 3.3 potentially affects all models of table 11.
\begin{table}[h!] 		
\centering
\begin{tabular}{|c|c|c|c|c|c|}
\hline
{\tt Field}&{\tt Spin}& {\tt Quantum Numbers}& {\tt Operator} & $C_1$& $C_3$\\
\hline
$A_\mu$& $1$& $(1,1,0)$& $\bar q'_{L}\gamma^\mu q'_{L}~ \bar \ell'_{L}\gamma_\mu \ell'_{L}$&$-1$& $0$\\ [2pt]
\hline	
$A^a_\mu$& $1$&$(1,3,0)$&$\bar q'_{L}\gamma^\mu \tau^a q'_{L}~\bar \ell'_{L}\gamma_\mu \tau^a \ell'_{L}$&$0$&$-1$\\  [2pt]
\hline
\hline
$U_\mu$& $1$&$(3,1,+2/3)$&$\bar q'_{L}\gamma^\mu \ell'_{L}~ \bar \ell'_{L}\gamma_\mu q'_{L}$&$-\frac{1}{2}$&$-\frac{1}{2}$\\  [2pt]
\hline
$U^a_\mu$& $1$&$(3,3,+2/3)$&$\bar q'_{L}\gamma^\mu\tau^a \ell'_{L}~ \bar \ell'_{L}\gamma_\mu \tau^aq'_{L}$&$-\frac{3}{2}$&$+\frac{1}{2}$\\  [2pt]	
\hline
$S$&$0$& $(3,1,-1/3)$&$\bar q'_{L} i\sigma^2{\ell'}^c_{L}~\overline{ i\sigma^2{\ell'}^c_{L}} q'_{L}$&$+\frac{1}{4}$&$-\frac{1}{4}$\\  [2pt]
\hline
$S^a$&$0$& $(3,3,-1/3)$&$\bar q'_{L}\tau^a i\sigma^2{\ell'}^c_{L}~\overline{ i\sigma^2{\ell'}^c_{L}}\tau^a q'_{L}$&$+\frac{3}{4}$&$+\frac{1}{4}$\\  [2pt]					
\hline
\end{tabular}
\caption{Spin zero and spin one mediators contributing, at tree-level, to the Lagrangian ${\cal L}_{NP}^0(\Lambda)$ of eq. (\ref{LNP}). 
Also shown are the operators they give rise to and the contribution to the coefficients $C_1$ and $C_3$ of the Lagrangian 
${\cal L}_{NP}^0(\Lambda)$, when a single fermion generation is involved.}
\end{table}

Assumption (ii) plays an important role in our analysis since it allows to strictly correlate the one-loop induced LFV and LFUV processes to the 
anomalous $B$ decays. In many specific SM extensions invoked to solve the $B$-anomalies NP mainly affects the third quark generation,
but our analysis shows that the near alignment between the primed basis and the quark mass basis is an additional very relevant condition.
To better illustrate this point, we can consider a slightly more general starting point, described by the NP effective Lagrangian:
\begin{equation}
{\cal L}_{NP}^0(\Lambda)=\frac{1}{\Lambda^2}
\left(
C_1~ \bar q'_{L}{\rm Q}\gamma^\mu q'_{L}~ \bar \ell'_{L}{\rm L}\gamma_\mu \ell'_{L}+C_3~\bar q'_{L}{\rm Q}\gamma^\mu \tau^a q'_{L}~ 
\bar \ell'_{L}{\rm L}\gamma_\mu \tau^a \ell'_{L}
\right)~,
\label{moreg}
\end{equation}
where Q and L are hermitian matrices in flavour space and generation indices are understood. The previous results are recovered when ${\rm Q}={\rm L}={\rm diag}(0,0,1)$. 
Concerning the lepton flavour matrix L, for simplicity we still choose ${\rm L}={\rm diag}(0,0,1)$, but we observe that the LFUV phenomenology is controlled by the diagonal entries of $L^e=V_e^\dagger L V_e$ and 
can be equally parametrized even considering a more general form of L. Such a general form would instead affect LFV processes, with predictions 
departing from our results.

Without loss of generality Q can be assumed diagonal in the primed basis. In this more general setup our results are still valid to a very good approximation,
provided the following two conditions are fulfilled: 
\begin{itemize}
\item[1.] The matrix Q has a non-vanishing trace and $|Q_{11}|\le |Q_{22}|\le |Q_{33}|$.
\item[2.] 
The primed basis is approximately aligned with the quark mass basis, that is the unitary transformations $V_{u,d}$ are close to the identity.
\end{itemize}
These two conditions extend the assumption (ii) by relaxing the requirement that NP dominantly affects the third generation.
It is not restrictive to assume assume ${\tt tr}({\rm Q})=+1$, by absorbing the size and the sign of ${\tt tr}({\rm Q})$ into the Wilson coefficients $C_{1,3}$.
In this more general setup the deviations $\delta R^{\mu /e}_{K}$ and $\delta R^{\tau/\ell}_{D^{(*)}}$ from the SM predictions are described by
\begin{eqnarray}
\delta R^{\mu /e}_{K} &\approx&  - \frac{2\pi}{\alpha |V_{ts}||C_9^{\mysmall\rm SM}|}\frac{v^2}{\Lambda^2} (C_1 + C_3) Q^d_{23}\lambda^e_{22}~~~,\nn\\
\delta R^{\tau/\ell}_{D^{(*)}} &\approx& - \frac{2v^2}{\Lambda^2} \,  C_3~ \left( \frac{V_{cd}}{V_{cb}} Q^d_{13}+\frac{V_{cs}}{V_{cb}} Q^d_{23} + Q^d_{33}\right)\lambda^{e}_{33}~~~,
\end{eqnarray}
where, in analogy with $\lambda^d$, we have defined $Q^d=V_d^\dagger Q V_d$.
Loop effects proportional to the gauge coupling constants $g_{1,2}^2$ are unchanged, since they are controlled by ${\tt tr}({\rm Q})$. Loop effects 
proportional to $y_t^2$ are obtained by the replacement:
\be
y_t^2 \lambda^u_{33}~~\to~~ y_t^2 (V_{CKM} Q^d V_{CKM}^\dagger)_{33}~~~.
\ee
Conditions 1. and  2. guarantees that the NP contributions to $R^{\mu /e}_{K}$ and $R^{\tau/\ell}_{D^{(*)}}$ are of the same order. In particular, $\delta R^{\mu /e}_{K}$ 
is proportional to $|Q^d_{23}|$ while, for sufficiently small $(V_d)_{ij}$ $(i\ne j)$ the deviation
$\delta R^{\tau/\ell}_{D^{(*)}}$ becomes proportional to $|Q^d_{33}|\gg |Q^d_{23}|$, thus allowing to overcome the enhancement $1/\alpha$ in $\delta R^{\mu /e}_{K}$.
Moreover in this regime we have $Q^d_{33} \approx (V_{CKM} Q^d V_{CKM}^\dagger)_{33}$, which
supplies the strict link between $R^{\tau/\ell}_{D^{(*)}}$ and the loop effects that are the object of our analysis. Notice that, provided $V_d$ is sufficiently close to the identity matrix,
the elements of $Q$ can also be of the same order. For instance a nearly degenerate matrix Q, arising for instance by an approximate $SU(3)$ flavor symmetry in the quark sector,
would still give rise to the loop effects discussed above. Of course for non-vanishing $Q_{11}$ and $Q_{22}$ we should carefully check whether
the constraints coming from processes involving quarks of the first two generations are satisfied.

The assumption that NP mainly affects the third quark generation can be relevant in NP scenarios
where the Lagrangian  (\ref{LNP}) arises from LQ exchange. Indeed, when integrating out LQ, we would get contact interactions 
of the type
\bea
&\lambda^{lq \,\dagger}_{jm} \lambda^{lq}_{ki} \, 
			(\bar q_{L i}\gamma^\mu X^a q_{L j})~ (\bar \ell_{L k}\gamma_\mu X^a\ell_{L m})&~~~~~{\rm (spin~0~~ LQ)\nn}\\
&\lambda^{lq \,\dagger}_{im} \lambda^{lq}_{kj} \, 
			(\bar q_{L i}\gamma^\mu X^a q_{L j})~ (\bar \ell_{L k}\gamma_\mu X^a\ell_{L m})&~~~~~{\rm (spin~1~~ LQ)}			
\label{lqflav}
\eea
where $X^a = (\mathds{1}, \tau^a)$ and the flavour structure is described by the matrix $\lambda^{lq}$. If  $\lambda^{lq}$ has rank 1 it can be decomposed 
as $\lambda^{lq}_{ij} = \theta^\ell_i \theta^q_j$. In this case we can always define $\lambda^{lq \,\dagger}_{im} \lambda^{lq}_{kj}=\lambda^{q}_{ij} \lambda^{l}_{km}$
with $\lambda^{l,q}$ matrices of rank 1 and we recover the flavour structure assumed in this paper.

Finally it is interesting to note that the assumption (ii) above is approximately realized in a class of models where  
the matrices $\lambda^f$ have the following pattern:
\begin{align}
\lambda^f = \frac{1}{c} \cdot \Delta_f \, Y_f \, \Delta_f^\dagger 
~~~~~~~~~~~~~~~~~~~~~~~~~~
\Delta_f = \left(
				\begin{array}{c c c}
						\alpha_f & & \\
						& \beta_f & \\
						& & 1
				\end{array}
				\right)~~~~,
\label{parametrization2}
\end{align}
where $c={\tt tr}(\Delta_f \, Y_f \, \Delta_f^\dagger)$,  $|\alpha| \ll |\beta| \ll 1$ and $Y_f$ are real matrices with entries of order one: $(Y_f)_{ij} \sim \cO(1)$.
For $\alpha_f\ll \beta_f\ll 1$ the eigenvalues of $\lambda_f$ are approximately 1, $\beta_f^2$ and $\alpha_f^2$ and the unitary matrices needed to
diagonalize $\lambda_f$ are close to the identity. 
The matrices $\lambda_f$ of our setup, eq.~(\ref{parametrization}), are recovered by taking
\begin{equation}
Y_f = \left(
				\begin{array}{c c c}
						1 & 1 & 1 \\
						1 & 1 & 1 \\
						1 & 1 & 1
				\end{array}
				\right) ~.
\label{parametrization3}
\end{equation}
Matrices $\lambda_f$ of the type in eq. (\ref{parametrization2}) typically arise in models with flavour symmetries,
such as abelian Froggatt-Nielsen symmetries or non-abelian $U(2)$ symmetries, or in models where fermion masses are generated through the mechanism
of partial compositeness. In all these scenarios, our conclusions still holds semi-quantitatively, i.e.~up to $\cO(1)$ corrections and barring accidental cancellations.

In the remaining of this section, we will briefly discuss different popular NP scenarios to explicitly analyse to what extent our 
assumptions hold and our conclusions can be applied.

\subsection{Models with Minimal Flavour Violation}

In the Minimal Flavour Violation (MFV) framework~\cite{DAmbrosio:2002vsn}, one assumes that the SM Yukawa couplings are the 
only sources of flavour breaking. Within this scheme, the most relevant FCNC four-fermion operators in the quark sector are of the 
form $(V-A)\times (V-A)$ as they are sensitive to the top yukawa coupling. In particular, the expression of $Q^d$ in MFV reads:
\begin{equation}
Q^d_{ij} = ( a \, \mathds{1} + b\, Y_U^\dagger Y_U )_{ij}
				      \approx a~ \delta_{ij} + b\, y_t^2 \, (V_{CKM})^*_{3i} (V_{CKM})_{3j} ~,
\label{MFVparam1}
\end{equation}
where $a$ and $b$ are real parameters. If $a=0$, after appropriate rescaling of $C_{1,3}$, eq.~(\ref{MFVparam1}) is clearly equivalent to eq.~(\ref{parametrization}), 
with $\alpha_d \!\propto\!  (V_{CKM})^*_{31}$, $\beta_d\!\propto\! (V_{CKM})^*_{32}$. As discussed above
also the case $a\ne 0$ can fulfill the more general conditions 1. and 2., provided $|a|\le |a+b y_t^2|$.   
The implementation of MFV in the lepton sector is not unique due to the ambiguity related to the neutrino masses. 
Assuming massless neutrinos, it turns out that   
\begin{equation}
L^e_{ij}  \approx (a' + b'\, y_{\ell_i}^2) \, \delta_{ij} \,.
\label{MFVparam2}
\end{equation}
Even though lepton flavor is conserved, LFUV is generated by the lepton Yukawa couplings $y_{\ell_i}$.
By assuming that $a'=0$, $L^e_{ij}$ has approximately rank 1.
The relative size of LFUV in the $\mu/e$ and $\tau/\mu$ sectors is of order $(m_\mu/m_\tau)^2$ 
which roughly corresponds to a loop factor. Therefore, in principle, MFV seems to be a promising 
setup where to accommodate simultaneously the $\RD$ and $\RKss$ anomalies~\cite{Alonso:2015sja}.
However, this possibility is challenged by the quantum effects discussed in \cite{Feruglio:2016gvd}  
and in the present paper.

\subsection[Models with \texorpdfstring{$U(2)$}{U(2)} flavour symmetries]{Models with \texorpdfstring{$\boldsymbol{U(2)}$}{U(2)} 
flavour symmetries}

Many explicit models proposed to address the $B$ anomalies, involving left-handed vector interactions  mediated by e.g.~$Z'$ 
or leptoquarks, rely on a $U(2)$ flavour breaking pattern~\cite{Barbieri:2015yvd,Barbieri:2016las,Bordone:2017anc}.
Common feature of models with $U(2)$ flavour symmetries is the existence of small flavour symmetry breaking parameters 
which suppress the coupling of NP with the first two fermion generations. We will collectively call $\epsilon$ 
such small $U(2)$-breaking terms.

In the case of a color-singlet mediator we can easily see that, after suitable rescaling, $Q$ and $L$ take the form~
\cite{Barbieri:2015yvd,Barbieri:2016las,Bordone:2017anc}:
\begin{align}
\left(
\begin{array}{c | c}
\phantom{\bigg(} 0 \phantom{\bigg)}     &     \epsilon \, \vec a_{Q,L}    \\
\hline
\epsilon \, \vec a_{Q,L}^\dagger     & 1
\end{array}
\right)
+ \cO(\epsilon^2) \,,
\label{LQmatrix}
\end{align}
where $\vec a_{Q,L}$ are generic $\cO(1)$ spurions of the underlying model.\footnote{Again, we have assumed that no trivial spurion $\mathds{1}$ arises.} 
For LQ mediators, we should first show that the flavour structure of the LQ model can be recast as in eq. (\ref{moreg}). As discussed above, a special case 
where this is granted is when $\lambda^{lq}$ is rank 1, see eq. (\ref{lqflav}). This is actually the case, since $\lambda^{lq}$ assumes the same form as in eq.~(\ref{LQmatrix}) 
and thus has rank 1 (up to corrections $\cO(\epsilon^2)$). In conclusion, our analysis holds at least semi-quantitatively in scenarios with $U(2)$ 
flavour symmetries.
  
\section{Discussion}  
Our results apply to the wide class of models where NP at the TeV scale mainly affects left-handed currents and the third generation.
In this framework a simultaneous explanation of both the $R_{D^{(*)}}^{\tau / \ell}$ and  $\RKss$ anomalies is strongly disfavoured. 
Therefore, it is important to examine in more detail the assumptions and the approximations made in our analysis.
At the same time it is of great interest to look for possible NP scenarios where our conclusions do not necessarily apply.

First of all our tools do not allow to fully reproduce the predictions of a UV complete theory giving rise to the semileptonic operators considered here.
In such a theory we expect finite contributions to the observables of interest, in particular the leptonic $Z$ coupling constants and the $\tau$ LFUV/LFV decays, beyond the leading logarithmic corrections explicitly computed here. Depending on the specific model, significant cancellations 
of the most constraining effects may take place when summing finite and logarithmic contributions. In such a case, our conclusions would be softened or even invalidated. Such finite contributions can originate already at the NP scale $\Lambda$. This is why in our analysis we have not included the finite corrections originating from the matching at the various thresholds: there is an intrinsic uncertainty due to the ignorance of the UV finite corrections and the knowledge of the full theory would be required to remove it.

A second remark concerns the set of semileptonic operators relevant to B anomalies that can be generated at the scale $\Lambda$ by a UV 
complete theory. While the choice of operators with only left-handed fermions is one of the preferred ones by the most recent fits, this does not exclude a leading or subleasing role of other operators. It is clear that a set of operators at the scale $\Lambda$ different from the one considered here in ${\cal L}^0_{NP}$ may lead to different conclusions. This scenario would deserve a new analysis, which goes beyond the scope of the 
present paper.

Moreover, also our assumption about the dominance of the third generation plays an important role. As we have seen in the previous section, many motivated flavour patterns predict such a dominance. However a completely different flavour pattern of semileptonic operators cannot be a priori excluded. For instance, NP could mainly give rise to operators selectively involving the bilinear $\bar q'_{3L}\gamma^\mu q'_{2L}$ and its conjugate.
This would contribute to both CC and NC anomalies, while evading the bounds from quantum effects.

Finally, even within the assumptions and the approximations made here, both the CC and the NC anomalies can be individually explained in our framework. It is only their simultaneous explanation that appear strongly disfavored. For instance the experimental value of $\RKss$ can be reproduced by choosing $\lambda_{23}^d\lambda^e_{22}\approx 1$ and $\Lambda/\sqrt{C_i}\approx 30~TeV$. In this case $R_{D^{(*)}}^{\tau / \ell}$ cannot significantly deviate from the SM prediction, but the RGE effects are negligible. Conversely, by taking 
$\lambda_{23}^d\lambda^e_{33}\approx 1$ and $\Lambda/\sqrt{C_3}\approx 5~TeV$ we can fit $R_{D^{(*)}}^{\tau / \ell}$ and decouple the loop effects.

\section{Conclusions}

The growing experimental indication of LFUV both in charged- and neutral-current semileptonic B-decays,
could represent the first indirect signal of New Physics. Since the required amount of LFUV is quite large,
one would expect that other NP signals should appear in other low- and/or high-energy observables. 
Indeed, in the recent literature, many studies focused on the experimental signatures implied by the solution of these anomalies
in specific scenarios, including kaon observables, kinematic distributions in $B$ decays, the lifetime of the $B_c^-$ meson, 
$\Upsilon$ and $\psi$ leptonic decays, tau lepton and dark matter searches.

In particular, in Ref.~\cite{Feruglio:2016gvd}, the importance of electroweak corrections for B anomalies has been highlighted,
assuming a class of gauge invariant semileptonic operators at the NP scale $\Lambda \gg v$, a common premise in many 
attempts to explain the B-anomalies.
The most important quantum effects turned out to be the modifications of the leptonic couplings of the $W$ and $Z$ vector bosons 
as well as the generation of a purely leptonic effective Lagrangian. It was found that the tight experimental bounds on $Z$-pole observables 
and $\tau$ decays challenge an explanation of  the LFUV observed in the charged and neutral-current channels.

In this work, we provided a detailed derivation of the relevant low-energy effective Lagrangian,
extending and generalizing the results of~\cite{Feruglio:2016gvd}. 
After defining the effective Lagrangian at the NP scale $\Lambda\sim 1$ TeV, we have discussed the required procedure of 
running and matching in order to determine the effective Lagrangian at the electroweak and lower scales. 
In particular, the running effects from $\Lambda$ down to the electroweak scale have been accounted for by the one-loop RGE 
in the limit of exact electroweak symmetry,
while from the electroweak scale down to the ${\rm GeV}$ scale by RGEs dominated by the electromagnetic interaction.
We have discussed the most relevant phenomenological implications of our setup focusing on $Z$-pole observables,
$\tau$ and $B$ meson decays. As a proof of correctness of our results, we have explicitly verified that the scale dependence of the 
RGE contributions from gauge and top Yukawa interactions cancels with that of the matrix elements in the relevant physical amplitudes. 
Finally, we have investigated the relevance of our results for specific classes of NP models such as minimal flavor violating models, 
$U(2)$ models and composite Higgs models. We find that, to good approximation, our conclusions apply to these models too, 
thus proving that the inclusion of electroweak quantum effects is mandatory for testing the consistency of the models with the existing data.
Our analysis shows that a simultaneous explanation of both the $R_{D^{(*)}}^{\tau / \ell}$ and  $\RKss$ anomalies is strongly disfavoured
in the adopted framework. However this conclusion should not be regarded as a no-go theorem. There are limitations in our approach
and fine-tuned solutions invoking model-dependent finite corrections that cannot be excluded. Moreover different conclusions can be expected when the present framework is generalized,
either by enlarging the set of initial operators or by allowing different flavour patterns. 
We believe that the most important message of
our work is that electroweak corrections should be carefully analysed in any framework where the explanation 
of B-anomalies invokes NP at the TeV scale.

\section*{Acknowledgements}

This work was supported in part by the MIUR-PRIN project 2010YJ2NYW and by the European Union network FP10 ITN ELUSIVES and INVISIBLES-PLUS
(H2020- MSCA- ITN- 2015-674896 and H2020- MSCA- RISE- 2015- 690575).
The research of P.P. is supported by the ERC Advanced Grant No.  267985 (DaMeSyFla), by the research grant TAsP, and by the INFN. 
The research of A.P. is supported by the Swiss National Science Foundation (SNF) under contract 200021-159720.

\end{document}